%% file: Cusp_4L_matter.tex
\documentclass[a4paper,11pt]{article}
\pdfoutput=1

\usepackage{jheppub}

\usepackage{hyperref}
\usepackage{graphicx}
\usepackage{amsmath}
\usepackage{amssymb}
\usepackage{xspace}
\usepackage{mathrsfs}
\usepackage{subfigure}
\usepackage{slashed}
\usepackage[normalem]{ulem}
\usepackage{xcolor} 
\usepackage{epstopdf}
\usepackage{soul}
\usepackage{booktabs}
\usepackage{enumitem}

\usepackage{bbm} 

\usepackage{tikz}
\usetikzlibrary{positioning,arrows}
\usepgflibrary{arrows.meta}
\usetikzlibrary{decorations.pathmorphing}
\usetikzlibrary{decorations.markings}
\usetikzlibrary{math}

\tikzset{
Wilson/.style={double distance=1.1pt,postaction={decorate}, decoration={markings,mark=at position .1 with {\arrow{Stealth[scale=1]}},mark=at position .98 with {\arrow{Stealth[scale=1]}}}},
Wilson_1/.style={double distance=1.1pt,postaction={decorate}, decoration={markings,mark=at position .08 with {\arrow{Stealth[scale=1]}},mark=at position .985 with {\arrow{Stealth[scale=1]}}}},
Wilson_3/.style={double distance=1.1pt,postaction={decorate}, decoration={markings,mark=at position .1 with {\arrow{Stealth[scale=1]}}}},
Wilson_Prop/.style={double distance=1.1pt,postaction={decorate}, decoration={markings,mark=at position .2 with {\arrow{Stealth[scale=1]}}}},
Wilson_blank/.style={double distance=1.1pt},
Wilson_arrow/.style={double distance=1.1pt,postaction={decorate}, decoration={markings,mark=at position .7 with {\arrow{Stealth[scale=1]}}}},
Scalar/.style={dashed},
Gluon/.style={decorate, draw=black, decoration={coil,aspect=0.5, post length = 2pt, pre length = 2pt, segment length=3pt,amplitude=3pt}},
Gluon_1/.style={decorate, draw=black, decoration={coil,aspect=0.5, post length = 0pt, pre length = 0pt, segment length=1.75pt,amplitude=1.5pt}},
Gluon_Vertex/.style={decorate, draw=black, decoration={coil,aspect=0.5, post length = 0pt, pre length = 0pt, segment length=3pt,amplitude=3pt}},
Incoming/.style={dashed, postaction={decorate}, decoration={markings,mark=at position 0.6 with {\arrow{Stealth[scale=1.4,reversed]}}}},
Photon/.style={decorate, draw=red, decoration={snake,segment length=3pt, amplitude=1.5pt}},
QuadProp/.style={}
 }

\tikzset{
    halfarrow/.style={postaction={decorate},
        decoration={markings,mark=at position .5 with
       {\arrow{Stealth[scale=1.2]}}}}}
       
\tikzset{axes_style/.style={->}} 
\tikzset{path_style/.style={halfarrow,blue}} 
\tikzset{help_lines_style/.style={dashed}} 
\tikzset{ana_strct_style/.style={dotted}} 
\tikzset{tick_style/.style={}} 
\tikzset{pointer_style/.style={>={Stealth}}} 

\DeclareRobustCommand{\eq}[1]{eq.~\eqref{eq:#1}}
\DeclareRobustCommand{\eqs}[2]{eqs.~\eqref{eq:#1} and \eqref{eq:#2}}
\DeclareRobustCommand{\eqss}[2]{eqs.~\eqref{eq:#1}\,--\,\eqref{eq:#2}}
\DeclareRobustCommand{\fig}[1]{figure~\ref{fig:#1}}

\DeclareRobustCommand{\sec}[1]{section~\ref{sec:#1}}

\DeclareRobustCommand{\app}[1]{appendix~\ref{app:#1}}

\DeclareRobustCommand{\tab}[1]{table~\ref{tab:#1}}

\usepackage{marginnote}

\newcommand{\ord}[1]{\mathcal{O}\!\left(#1\right)}

\newcommand{\df}{\mathrm{d}}

\newcommand{\eps}{\epsilon}

\newcommand{\nn}{\nonumber}

\newcommand{\cusp}{\mathrm{cusp}}

\newcommand{\one}{{(1)}}

\newcommand\lr[1]{{\left({#1}\right)}}

\newcommand\TabRef{[$\times$]}
\newcommand{\CS}[2]{#1^{#2}}
\newcommand{\CSf}[2]{#1^{f\! #2}}
\newcommand{\CSff}[2]{#1^{f\!f\! #2}}
\newcommand{\CSfff}[2]{#1^{f\!f\!f\! #2}}

\newcommand{\tagaligneq}{\refstepcounter{equation}\tag{\theequation}}

\def\df{\textrm{d}}

\allowdisplaybreaks[2]

\preprint{\begin{flushright}
MITP/18-125\\
MPP-2019-1
\end{flushright}}

\title{
Matter dependence of the four-loop QCD cusp anomalous dimension: from small angles to all angles
}

\author[a]{Robin Br\"user,}
\author[a,b,c]{Andrey Grozin,}
\author[d]{Johannes M. Henn,}
\author[a]{Maximilian Stahlhofen}

\affiliation[a]{PRISMA Cluster of Excellence, Johannes Gutenberg University, 55128 Mainz, Germany}
\affiliation[b]{Budker Institute of Nuclear Physics SB RAS, Novosibirsk 630090, Russia}
\affiliation[c]{Novosibirsk State University, Novosibirsk 630090, Russia}
\affiliation[d]{Max-Planck-Institut f\"ur Physik, Werner-Heisenberg-Institut, 80805 M\"unchen, Germany}

\emailAdd{brueser@uni-mainz.de}
\emailAdd{A.G.Grozin@inp.nsk.su}
\emailAdd{henn@mpp.mpg.de}
\emailAdd{mastahlh@uni-mainz.de}

\abstract{
We compute the fermionic contributions to the cusp anomalous dimension in QCD at four loops as an expansion for small cusp angle.
As a byproduct we also obtain the respective terms of the four-loop HQET wave function anomalous dimension.
Our new results at small angles provide stringent tests of a recent conjecture for the exact angle dependence of the matter terms in the four-loop cusp anomalous dimension.
We find that the conjecture does not hold for two of the seven fermionic color structures,
but passes all tests for the remaining terms.
This provides strong support for the validity of the corresponding conjectured expressions with full angle dependence.
Taking the limit of large Minkowskian angle, we extract novel analytic results for certain terms of the light-like cusp anomalous dimension.
They agree with the known numerical results. 
Finally, we study the anti-parallel lines limit of the cusp anomalous dimension. 
In a conformal theory, the latter is proportional to the static quark-antiquark potential. 
We use the new four-loop results to determine parts of the conformal anomaly term.
}

\begin{document}
\maketitle

\section{Introduction}
\label{sec:Intro}

The cusp anomalous dimension is a universal and ubiquitous quantity in QCD and the effective field theories describing its IR behavior as e.g. heavy quark effective theory (HQET) and soft collinear effective theory (SCET). It governs the IR singularity structure of QCD scattering amplitudes \cite{Ivanov:1985np,Korchemsky:1985xu,Korchemsky:1985ts,Korchemsky:1985xj,Korchemsky:1986fj}. In the presence of massive partons the IR divergences are controlled by the angle dependent cusp anomalous dimension $\Gamma_\textrm{cusp}(\phi,\alpha_s)$. 
It can be determined from the UV divergences of a time-like  Wilson loop with a cusp of (Euclidean) angle $\phi$~\cite{Korchemsky:1985xj}. 
The light-like cusp anomalous dimension $K(\alpha_s)$ relevant for scattering of massless partons emerges as the $\phi \to i \infty$ limit of $\Gamma_\textrm{cusp}(\phi,\alpha_s)$ \cite{Korchemsky:1985xj,Korchemsky:1991zp}. 
It is the key ingredient to Sudakov resummation for scattering processes at high-energy colliders.

\begin{table}[ht]
\begin{center}
\begin{tabular}{llllp{2cm}l}
\toprule
color structure 		& sample diagram		& $\Gamma_{\text{cusp}}(\phi)$ 		& $\phi\ll1$ 				& light-like 							& $\gamma_h$ \\
 \midrule 
$(T_F n_f)^3 C_R $ 		& \input{Plots/Dia_fffR.tex} 	& \cite{Grozin:2004yc} 			& \cite{Grozin:2004yc}			& \cite{Beneke:1995pq}						& \cite{Broadhurst:1994se} \\	
\midrule
$(T_F n_f)^2 C_R C_F $ 		& \input{Plots/Dia_ffRF.tex} 	& \cite{Grozin:2015kna,Grozin:2016ydd} 	& \cite{Grozin:2015kna,Grozin:2016ydd} 	& \cite{Grozin:2015kna,Grozin:2016ydd}				& \cite{Grozin:2015kna,Grozin:2016ydd}  \\
$(T_F n_f)^2 C_R C_A $ 		& \input{Plots/Dia_ffRA.tex} 	& \TabRef 				& \TabRef 				& \cite{Henn:2016men,Davies:2016jie} 				& \cite{Marquard:2018rwx} \TabRef \\
\midrule
$(T_F n_f) C_R C_F^2$ 		& \input{Plots/Dia_fRFF.tex}	& \cite{Grozin:2018vdn} 		& \cite{Grozin:2018vdn}			& \cite{Grozin:2018vdn}						& \cite{Grozin:2018vdn} \\
$(T_F n_f) C_R C_F C_A$		& \input{Plots/Dia_fRFA.tex} 	& \TabRef				& \TabRef 				& \cite{Moch:2017uml}$^*$ \TabRef				& \cite{Marquard:2018rwx}$^*$ \TabRef \\
$(T_F n_f) C_R C_A^2 $ 		& \input{Plots/Dia_fRAA.tex} 	& 					& \TabRef 				& \cite{Moch:2017uml}$^*$ \TabRef				& \cite{Marquard:2018rwx}$^*$ \TabRef \\
$n_f \frac{d_R d_F}{N_R}$ 	& \input{Plots/Dia_fdRdF.tex}	& 					& \cite{Grozin:2017css} \TabRef 	&   \cite{Moch:2017uml,Moch:2018wjh}$^*$ \newline \cite{Lee:2019zop,Henn:2019rmi} & \cite{Grozin:2017css} \\
\rule{0pt}{2em} $n_f^1$, $N_c\to\infty$ 	&  		& 					& \TabRef 				& \cite{Henn:2016men,Moch:2017uml} 				& \TabRef  \\
\midrule
$C_R C_A^3$ 			& \input{Plots/Dia_RAAA.tex}	& 					& 					& \cite{Moch:2017uml,Moch:2018wjh}$^*$ 					& \cite{Marquard:2018rwx}$^*$ \\
$\frac{d_R d_A}{N_R}$ 		& \input{Plots/Dia_dRdA.tex}	& 					& 					& \cite{Moch:2017uml,Moch:2018wjh}$^*$ 				& \cite{Marquard:2018rwx}$^*$ \\
\rule{0pt}{2em}$n_f^0$, $N_c\to\infty$ 	&  			& 					&  					& \cite{Lee:2016ixa,Moch:2017uml} 				&  \\
\bottomrule
\end{tabular}
\end{center}
\caption{Four-loop contributions to $\Gamma_{\text{cusp}}(\phi)$ and its limits in QCD as well as the HQET field anomalous dimension $\gamma_h$.
The $^*$ marks numerical results. The results \TabRef~are obtained in the present paper.
}
\label{tab:status}
\end{table}

In addition to the light-like limit, there are two more interesting limits, the anti-parallel lines and the small angle limit. In the anti-parallel lines limit the cusp anomalous dimension is in one-to-one correspondence with the static quark-antiquark potential up to terms that are due to the conformal anomaly of (massless) QCD \cite{Kilian:1993nk,Correa:2012at,Grozin:2015kna}. At small cusp angle the cusp anomalous dimension is given by a regular expansion $\Gamma_\textrm{cusp}(\phi,\alpha_s)=-\phi^2 B(\alpha_s)+\mathcal{O}(\phi^4)$ in $\phi^2$, where $B(\alpha_s)$ is the Bremsstrahlung function \cite{Correa:2012at}. The Bremsstrahlung function describes the radiation loss of a slowly moving heavy quark in an external gauge field. Furthermore, the cusp anomalous dimension determines the renormalization group (RG) running of the Isgur-Wise function, a universal
function in HQET \cite{Isgur:1989vq,Isgur:1989ed}. At small cusp angles it is used for the extraction of the CKM matrix element $V_{cb}$ from semileptonic $B \rightarrow D^{(*)}$ decays, see e.g. \cite{Neubert:1993mb}.

The  QCD cusp anomalous dimension $\Gamma_\textrm{cusp}(\phi,\alpha_s)$  is known up to three loops \cite{Grozin:2014hna,Grozin:2015kna} in perturbation theory for arbitrary $\phi$. At four loops partial results are available and summarized in \tab{status}.
At four loops in  $\mathcal{N}=4$ super Yang Mills (sYM) theory the light-like limit of the cusp anomalous dimension has been computed numerically \cite{Boels:2017skl,Boels:2017ftb} and the full angle dependence is known analytically in the planar limit \cite{Henn:2013wfa}. In addition, the Bremsstrahlung function is known exactly (to all loop orders, and including the full color dependence) in  $\mathcal{N}=4$ sYM \cite{Correa:2012at}.

In \cite{Grozin:2014hna,Grozin:2015kna} it was observed that, up to three loops, the cusp anomalous dimension has a universal structure.
Namely, expanding $\Gamma_\textrm{cusp}(\phi)$ in $K(\alpha_s)$ instead of $\alpha_s$, the coefficients of $K^n(\alpha_s)$ are universal. In particular, they are equal in QCD, pure Yang-Mills, and $\mathcal{N}=4$ sYM.
Based on their observation the authors of \cite{Grozin:2014hna,Grozin:2015kna} conjectured that this universality holds to all orders in perturbation theory.
The conjecture allows to predict the fermionic contributions to $\Gamma_\cusp(\phi)$ in QCD at a given loop order (up to a normalization factor) using only lower loop results as an input. By `fermionic' contributions we refer to terms that depend on $n_f$, the number of active fermion flavors.

A major goal of this paper is to check the validity of these predictions at four loops.
The idea is to predict the fermionic four-loop terms with full angle dependence and verify the conjectured expressions against analytic results for $\Gamma_\cusp(\phi)$ calculated in the small angle expansion.
This analysis was initiated in \cite{Grozin:2017css} by investigating the $n_f$ term proportional to the quartic Casimir color factor.
It was found that the conjecture does not hold for that particular color structure.
This may be connected to the special nature of the quartic Casimir contributions, which appear at four loops for the first time in the perturbative expansion and are the reason for Casimir scaling violation.
It is therefore interesting to ask whether the conjecture possibly holds for the other four-loop color structures.
For the terms proportional to $(n_f T_F)^3C_R$, $(n_f T_F)^2C_R C_F$  and $(n_f T_F)C_R C_F^2$, in the following called `Abelian' color structures (as they are independent of $C_A$), a quick answer can be given:
these contributions are known exactly, cf. \tab{status}, and exactly comply with the conjecture.

This encouraged us to extend the analysis also to the other four-loop $n_f$ contributions, where no explicit all-angles result is available to date.
In the present paper we therefore compute the corresponding terms up to $\ord{\phi^4}$ and partly $\ord{\phi^6}$ in the small angle expansion and use them to test the conjecture.
From the $\ord{\phi^0}$ term of our Wilson loop calculation we obtain analytical expressions for the four-loop  heavy quark field anomalous dimension in heavy quark effective theory (HQET).
Our approach closely follows the lines of \cite{Grozin:2017css}.
In addition, we study the anti-parallel lines limit of the cusp anomalous dimension. Given our findings regarding the validity of the conjecture at four loops we derive new terms in its relation to the static quark-antiquark potential.

The paper is organized as follows. In \sec{Setup} we present the  setup and in  \sec{calculation}  details of our calculation. Section \ref{sec:results}  contains our four-loop results for the small angle expansion of the cusp anomalous dimension as well as the HQET field anomalous dimension. In \sec{conjecture} we review the conjecture of \cite{Grozin:2014hna,Grozin:2015kna}, test it against our results for the small angle expansion and discuss the outcome. Section \ref{sec:APL} elaborates on the consequences for the relation between the cusp anomalous dimension and the static potential. We conclude in \sec{conclusion}. 
In \app{conj_res} we collect all conjectured expressions for the four-loop cusp anomalous dimension with full angle dependence.

\section{Definitions and ultraviolet properties of Wilson line operators}
\label{sec:Setup}

\begin{figure}
\centering
 \input{Plots/Wilson_Cusp.tex}
 \caption{Wilson line with two straight line segments forming a cusp with Euclidean cusp angle $\phi$.}
 \label{fig:cusp}
\end{figure}
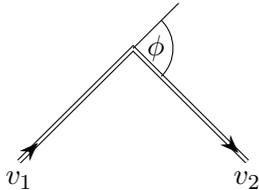

We start with the definition of the cusp anomalous dimension in QCD.
To this end we consider a closed Wilson loop with a time-like integration contour $C$
\begin{align} \label{eq:DefW}
W = \frac{1}{N_R} \textrm{Tr}_R \big\langle0|\, P\, \exp \lr{i g\oint_C dx^\mu 
A_\mu(x)}|0\big\rangle = 1+ \ord{g^2}\,.
\end{align}
Here $A_\mu=A_\mu^a\,T_R^a$ is the gluon field, $P$ is the path-ordering 
operator and the trace is over (color) indices in the representation $R$ of the 
gauge group $SU(N_c)$. A cusp in the integration contour gives rise to UV divergences, which are renormalized multiplicatively \cite{Korchemsky:1987wg}. The associated anomalous dimension depends on the cusp angle $\phi$ and is correspondingly called cusp anomalous dimension $\Gamma_\textrm{cusp}(\phi,\alpha_s)$. To compute it we conveniently consider a contour consisting of two straight line segments along directions $v_1^\mu$ and $v_2^\mu$ ($v_1^2=v_2^2=1$), which form a cusp in the origin and extend to infinity where the contour is closed, see \fig{cusp}. 
The Euclidean cusp angle is defined by $\cos \phi=v_1 \cdot v_2$. 
In Minkowskian spacetime and for real $v_1$ and $v_2$ the Euclidean cusp angle is purely imaginary leading to the definition of the Minkowskian cusp angle $\cosh \varphi=v_1 \cdot v_2$, such that $\phi=i \varphi$. The full angle-dependent cusp anomalous dimension was computed up to three loops in \cite{Polyakov:1980ca,Korchemsky:1987wg,Correa:2012nk,Grozin:2014hna,Grozin:2015kna} 
and we use the same setup as in \cite{Grozin:2015kna} to calculate it in the small angle expansion at four loops.

The cusp anomalous dimension appears also in the context of heavy quark effective theory (HQET), see e.\,g.~\cite{Neubert:1993mb,Manohar:2000dt,Grozin:2004yc}. In the HQET picture the Wilson line configuration depicted above corresponds to a heavy quark moving with four-velocity $v_1^\mu$, then scattering at an external (electromagnetic) source to instantaneously change its velocity to $v_2^\mu$. This allows us to compute the cusp anomalous dimension using HQET momentum space Feynman rules for the heavy quark propagator and the heavy-quark--gluon vertex
\begin{align}
& \begin{tikzpicture}[scale=1.7, baseline=(current bounding box.center)]	  
	  \draw[Wilson_Prop] (0,0)  --  node[pos=0.1, anchor = south] {$v$} (1.2,0)  {};	  
	  \draw[->] (0.4,-0.1)  --  node[pos=0.5, anchor = north] {$p$} (1,-0.1)  {};
\end{tikzpicture}= \frac{i}{v \cdot p +\delta} \; , & &
\begin{tikzpicture}[scale=1.7]	  
	  \draw[Wilson_Prop] (0,0)  --  node[pos=0.1, anchor = south] {$v$} (1.2,0)  {};
	  \draw[Gluon_Vertex] (0.6,0.5) -- (0.6,0) {};	  
\end{tikzpicture} = ig v^\mu T_R^a \; . & 
\end{align}
The double line represents a heavy quark (or Wilson) line with $v^2=1$ in the $SU(N_c)$ representation $R$. We take the heavy quark to be slightly off-shell $\delta \neq 0$ in order to regulate IR divergences in (\ref{eq:DefW}), since we are only interested in the UV divergences arising from loops involving the cusp. The off-shellness $\delta$ can be interpreted as the residual energy of the heavy quark.  Without loss of generality we choose $\delta=-1/2$ in our calculation.

We distinguish two types of Feynman diagrams contributing to the Wilson loop  in (\ref{eq:DefW}). The first are HQET self-energy diagrams, which are $\phi$-independent. The second are (one-particle-irreducible) vertex corrections depending on the cusp angle. The sum of the former is related to the sum of the latter at zero cusp angle by a HQET Ward identity. Denoting the sum of the vertex diagrams as $V(\phi)$, we thus have~\cite{Korchemsky:1987wg}
\begin{align}\label{eq:lnW}
\log W = \log V(\phi) - \log V(0) = \log Z + \ord{\eps^0}\,,
\end{align}
where we have introduced the $\overline{\text{MS}}$ cusp renormalization constant $Z$. Here and 
throughout this paper we use dimensional regularization with $d=4-2\eps$ and we expand as
\begin{align}
\Gamma_\cusp(\phi,\alpha_s) 
&= \sum_{k \ge 1} \left( \frac{\alpha_s}{\pi} \right)^k \Gamma_\text{cusp}^{(k)}(\phi)\,.
\end{align}
The cusp anomalous dimension is defined by the renormalization group equation (RGE)
\begin{align}\label{eq:cusp-def}
\Gamma_\cusp(\phi,\alpha_s) &= \frac{\df\log Z}{\df\log\mu} 
\;,
\end{align}
where $\mu$ is the renormalization scale. Iteratively solving this equation yields
\begin{align}\label{eq:LogZ}
\log Z 
=& -\frac{\alpha_s}{\pi}\frac{\Gamma_\textrm{cusp}^{(1)}}{2 \epsilon} 
+ \left(\frac{\alpha_s}{\pi} \right)^2 \left[ \frac{\beta_0 \Gamma_\textrm{cusp}^{(1)}}{16 \epsilon^2} - \frac{\Gamma_\textrm{cusp}^{(2)}}{4 \epsilon} \right] \\
&+ \left(\frac{\alpha_s}{\pi} \right)^3 \left[ -\frac{\beta_0^2 \Gamma_\textrm{cusp}^{(1)}}{96 \epsilon^3} +\frac{\beta_1 \Gamma_\textrm{cusp}^{(1)}+4\beta_0 \Gamma_\textrm{cusp}^{(2)}}{96 \epsilon^2} - \frac{\Gamma_\textrm{cusp}^{(3)}}{6 \epsilon} \right]
+ \left(\frac{\alpha_s}{\pi} \right)^4 \left[ 
+\frac{\beta_0^3 \Gamma_\textrm{cusp}^{(1)}}{512 \epsilon^4} \right.\nn\\
& \left. -\frac{\beta_0 \left(\beta_1 \Gamma_\textrm{cusp}^{(1)}+2\beta_0 \Gamma_\textrm{cusp}^{(2)} \right)}{256 \epsilon^3} 
+\frac{\beta_2 \Gamma_\textrm{cusp}^{(1)}+4 \beta_1 \Gamma_\textrm{cusp}^{(2)}+16\beta_0 \Gamma_\textrm{cusp}^{(3)}}{512 \epsilon^2} 
- \frac{\Gamma_\textrm{cusp}^{(4)}}{8 \epsilon} \right]
+\mathcal{O}(\alpha_s^5) \nn
\,.
\end{align}
From this expression we see that, at a given order in perturbation theory, the poles higher than $1/\epsilon$ are determined by lower order results of the cusp anomalous dimension and the $\beta$-function of QCD. The only new information at each loop enters through the $1/\epsilon$ term. For the $\beta$-function we use
\begin{align}
\frac{\df \log \alpha_s}{\df \log \mu}= -2 \epsilon+2 \beta(\alpha_s)\;, \qquad \beta(\alpha_s)=- \sum_{k \ge 0} \left(\frac{\alpha_s}{4 \pi} \right)^{k+1} \beta_k \,,
\end{align}
with $\beta_0 = \frac{11}{3}\,C_A -\frac{4}{3}\,T_F\,n_f$. 
The relevant higher order coefficients can e.g. be found in \cite{Chetyrkin:2017bjc,Luthe:2017ttg}. 
The parameter $n_f$ denotes the number of active fermion flavors.

So far we considered the angle dependent cusp anomalous dimension $\Gamma_\textrm{cusp}(\phi,\alpha_s)$, originating from the UV divergences of a cusped Wilson loop with a time-like integration contour. For a light-like integration contour the corresponding light-like cusp anomalous dimension is denoted by $K(\alpha_s)$. The light-like limit is reached for $\phi \rightarrow i \infty$ or equivalently for $\varphi \rightarrow \infty$ and we have \cite{Korchemsky:1985xj,Korchemsky:1991zp}
\begin{align}
 \Gamma_\textrm{cusp}(\phi,\alpha_s) = - i \phi \,K(\alpha_s) \quad \textrm{for }\phi \rightarrow i \infty \, .
\end{align}

Finally we briefly discuss the heavy quark field renormalization in HQET. The corresponding $\overline{\text{MS}}$ renormalization constant $Z_h$ can be obtained from the derivative of the HQET self energy $\Sigma_h(\delta)$ w.r.t. the residual energy $\delta$ of the heavy quark using
\begin{align}
 \log\left(1 - \frac{\df\Sigma_h}{\df\delta}\right) = - \log Z_h + \mathcal{O}(\epsilon^0)\,.
 \label{eq:Zh1}
\end{align}
Note that the $L$-loop term of $\Sigma_h(\delta)$ is proportional to $\delta^{1-2L\epsilon}$ according to dimensional analysis. Using the HQET Ward identity
\begin{align}
V(0) = 1 - \frac{\df\Sigma_h}{\df\delta}\,,
\label{eq:Ward}
\end{align}
we can relate the renormalization constant to the vertex function at zero cusp angle
\begin{align}
\log V(0) = - \log Z_h + \mathcal{O}(\epsilon^0)\,.
\label{eq:Zh2}
\end{align}
This quantity is not gauge invariant (and not observable).
The corresponding HQET heavy quark field anomalous dimension therefore also depends not only on the strong coupling $\alpha_s$, but also on the gauge. Choosing generalized covariant gauge a dependence on the gauge parameter $\xi$ remains:
\begin{align}
\gamma_h(\alpha_s,\xi) = \frac{\df \log Z_h}{\df \log \mu}\,.
\label{eq:gammadef}
\end{align}
This RGE can be solved analogously to \eq{LogZ} in a perturbative fashion. This time, however, we have to take the dependence on the gauge parameter $\xi(\mu)$ into account, which itself is renormalization scale dependent. The relevant terms of the corresponding anomalous dimension can e.g. be found in \cite{Chetyrkin:2017bjc}.

\section{Four-loop calculation of matter-dependent terms at small angle}
\label{sec:calculation}

In this section we describe the computation of the cusp anomalous dimension at four loops in the small angle expansion. We begin with the discussion of the color structures. Then we describe the general computational workflow, including the calculation of the Feynman diagrams, partial fraction decomposition, integration-by-parts reduction and the computation of the master integrals. Here we follow in the most parts \cite{Grozin:2017css}. Finally we point out a subtlety in the renormalization procedure of the off-shell and therefore gauge-dependent Wilson loop in order to obtain \eq{LogZ}.

\subsection{Color dependence of $\Gamma_\cusp$ to four loops}
The structure of the QCD cusp anomalous dimension in terms of color factors is determined by non-Abelian exponentiation \cite{Gatheral:1983cz,Frenkel:1984pz,Sterman:1981jc}. Up to three loops we have
\begin{align}
 \Gamma_\textrm{cusp}={}& \frac{\alpha_s}{\pi} C_R \CS{\Gamma_\text{cusp}}{R}
+ \left(\frac{\alpha_s}{\pi} \right)^2 C_R \left[ \CS{\Gamma_\text{cusp}}{RA} + n_f T_F \CSf{\Gamma_\text{cusp}}{R} \right]
+ \left(\frac{\alpha_s}{\pi} \right)^3 C_R \Big[ C_A^2 \CS{\Gamma_\text{cusp}}{RAA}  \nn \\
&+ (n_f T_F)^2 \CSff{\Gamma_\text{cusp}}{R} 
+n_f T_F \left(C_F \CSf{\Gamma_\text{cusp}}{RF}+ C_A \CSf{\Gamma_\text{cusp}}{RA} \right) \Big] 
+ \mathcal{O}(\alpha_s^4) \; . \label{eq:Color123L}
 \end{align}
 The quadratic Casimir operators $C_R$, $C_A$, $C_F$ are defined according to $T_R^aT_R^a=C_R \mathbbm{1}_R$, where $T_R^a$ is the generator of the $SU(N_c)$ representation $R$ with $\textrm{Tr}[T_R^a T_R^b]=T_R \delta^{ab}$ and $\textrm{Tr}[ \mathbbm{1}_R]=N_R$. In QCD the two relevant representations are the adjoint ($R=A$) and the fundamental representation ($R=F$) with $T_F=1/2$. Up to three loops the cusp anomalous dimension obeys Casimir scaling, i.e. it depends linearly on $C_R$, where $R$ is the representation of the Wilson loop. 
Starting at four loops, however, Casimir scaling is violated by terms proportional to  quartic Casimir operators. There are two types of such color factors at this order, which we denote by $d_R d_A/N_R$ and $n_f d_R d_F/N_R$. Like the color factor $C_R C_A^3$ the former belongs to the purely gluonic part of the cusp anomalous dimension.
The quartic Casimir operators are defined by symmetrized traces\footnote{
The symmetrized trace is defined by $\textrm{STr}[T^{a_1} \cdots T^{a_n}]=\frac{1}{n!} \sum_{\sigma \in S_n}\textrm{Tr}[T^{a_{\sigma(1)}} \cdots T^{a_{\sigma(n)}}]$, where the sum runs over all permutations of indices.}
of four generators \cite{vanRitbergen:1998pn}
\begin{align} \label{eq:def_quartic}
 \frac{d_R d_{R^\prime}}{N_R} \equiv \frac{d_R^{abcd} d_{R^\prime}^{abcd}}{N_R} \;, \qquad d_{R}^{abcd}= \textrm{STr}\left[ T_{R}^a T_{R}^b T_{R}^c T_{R}^d \right]  \; .
\end{align}
At four loops $\Gamma_\cusp$ then takes the form
\begin{align}
 \Gamma_\text{cusp}^{(4)} =& \; C_R C_A^3 \CS{\Gamma_\text{cusp}}{RAAA}+\frac{d_R d_A}{N_R} \CS{\Gamma_\text{cusp}}{dRA}+  (n_f T_F)^3 C_R \CSfff{\Gamma_\text{cusp}}{R} + (n_f T_F)^2  \! \left( C_R C_F \CSff{\Gamma_\text{cusp}}{RF} +C_R C_A \CSff{\Gamma_\text{cusp}}{RA} \right) \nn \\
 &+ n_f T_F  \left( C_R C_F^2 \CSf{\Gamma_\text{cusp}}{RFF} +C_R C_F C_A \CSf{\Gamma_\text{cusp}}{RFA} +C_R C_A^2 \CSf{\Gamma_\text{cusp}}{RAA}  \right) + n_f \frac{d_R d_F}{N_R} \CSf{\Gamma_\text{cusp}}{dRF} \;.
 \label{eq:Color4L}
\end{align}
In \tab{status} we show for each of the four-loop color structures an example of a contributing Feynman diagram. We will employ the analogous notation regarding the coefficients of the different color factors in \eqs{Color123L}{Color4L} also for the light-like cusp anomalous dimension $K$ and the HQET field anomalous dimension $\gamma_h$.

The fermionic quartic Casimir operator $n_f d_R d_F/N_R$ only appears in 18 Feynman diagrams at four loops. Those are the diagrams with a fermion box subdiagram, where the four gluons are directly attached to the Wilson lines as shown in the (planar) sample diagram in \tab{status}. The gluonic quartic Casimir operator does not only appear in the corresponding diagrams with a gluon or ghost box, but for instance also in diagrams like the example diagram for $d_R d_A/N_R$ shown in \tab{status}. That diagram has the color factor
\begin{align}
 \frac{1}{N_R}\textrm{Tr} \left[T_{R}^a T_{R}^b T_{R}^c T_{R}^d  T_{R}^a T_{R}^b T_{R}^c T_{R}^d \right]
 = \frac{1}{N_R}\textrm{Tr} \left[T_{R}^a T_{R}^b T_{R}^c T_{R}^d \right] \textrm{Tr} \left[ T_{A}^a T_{A}^b T_{A}^c T_{A}^d \right] + \dots = \frac{d_R d_A}{N_R} + \dots \,,
\end{align}
where the ellipses denote terms involving only quadratic Casimir operators and in the second step we repeatedly used the Lie algebra $[T_R^a, T_R^b]=i f^{abc} T_R^c$ and $\big(T_{A}^b\big)_{ac} = i f^{abc}$. In a similar way also diagrams with three gluon vertices give rise  to $d_R d_A/N_R$ terms. This illustrates that the gluonic quartic Casimir operator appears in a much larger set of Feynman diagrams than $n_f d_R d_F/N_R$.

\subsection{Calculation of Feynman diagrams}
Next we outline our calculation of the four-loop cusp anomalous dimension expanded in small cusp angle $\phi$. More details on the calculation, especially on the analytic evaluation of the master integrals, can be found in \cite{Grozin:2017css}, where the quartic Casimir $n_f d_R d_F/N_R$ contribution was studied. Since we are dealing with many more Feynman diagrams than \cite{Grozin:2017css}, we automatize the calculation to a higher degree. The Feynman diagrams are generated with {\tt qgraf}~\cite{Nogueira:1991ex}, then mapped to integral topologies and after this the color, Dirac and Lorentz algebra is performed using a dedicated Mathematica code. 

The integrals appearing in the Feynman diagrams are regular at $\phi=0$. To obtain their small angle expansion we can therefore simply expand their integrand in a Taylor series. In this way we get a power series in $\phi$, where the coefficients are given by linear combinations of tensor integrals. We perform a tensor reduction to relate the tensor integrals to scalar integrals. After that only even powers of $\phi$ survive upon integration, since the original integrals before expansion are functions of $\cos \phi$. We end up with expressions for each diagram in terms of scalar four-loop integrals of (Wilson line) propagator-type.  

Depending on the Feynman diagram, it may contain integrals with linearly dependent HQET-type propagators. For example the sample diagrams for the color structures $d_R d_A/N_R$ and $n_f T_F C_R C_A^2$ in \tab{status} give rise to such integrals in the small angle expansion. In order to prepare them for straightforward integration-by-part (IBP) reduction, we remove the linear dependences between propagators beforehand. This is achieved by applying the multivariate partial fraction decomposition algorithm outlined in \cite{Pak:2011xt}. With the help of a Gr\"obner basis, the algorithm constructs for a given integral topology with linearly dependent propagators a set of replacement rules. 
For any integral belonging to that topology we then obtain an appropriate partial fraction decomposition by simply applying these rules recursively. 
The use of a Gr\"obner basis ensures that the recursion terminates.
The result is a sum of integrands with linearly independent propagators.

\subsection{Integral topologies and master integrals}
\label{sec:Integrals_Topos}
\begin{figure}[t]
\centering
\begin{minipage}{0.31\textwidth}
\centering
\input{Plots/Topo_1_box.tex}
\end{minipage}
\begin{minipage}{0.31\textwidth}
\centering
\input{Plots/Topo_2_box.tex}
\end{minipage}
\begin{minipage}{0.31\textwidth}
\centering
\input{Plots/Topo_3_box.tex}
\end{minipage}
\begin{minipage}{0.48\textwidth}
\centering
\input{Plots/Topo_4.tex}
\end{minipage}
\begin{minipage}{0.45\textwidth}
\centering
\input{Plots/Topo_5.tex}
\end{minipage}

\caption{Integral topologies of the master integrals of all $n_f$ dependent color structures of the cusp anomalous dimension in the small angle expansion.}
\label{fig:integral_topos}
\end{figure}
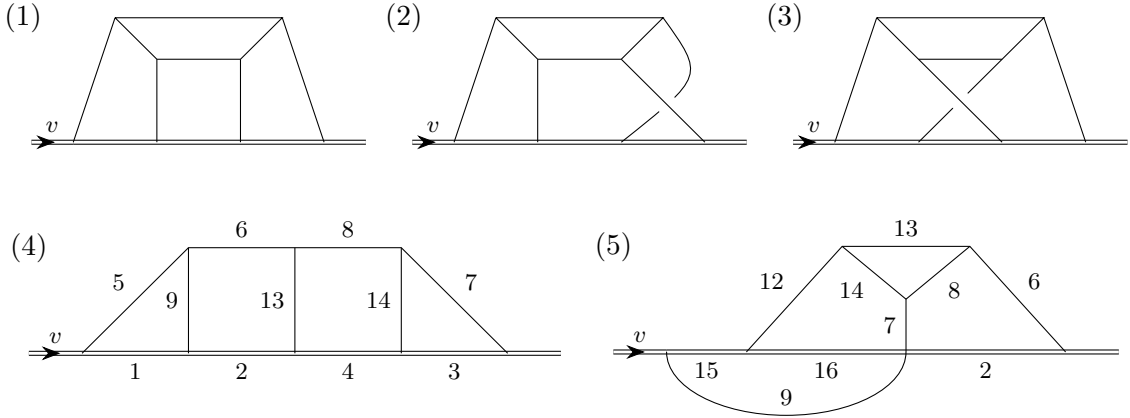

For the IBP reduction we use {\tt FIRE5}~\cite{Smirnov:2014hma} in combination with {\tt LiteRed}~\cite{Lee:2012cn,Lee:2013mka}.  After the IBP reduction we are left with 46 master integrals (MI), belonging to five integral topologies shown in figure \ref{fig:integral_topos}. The integral families associated with the first three topologies reduce to 43 MI, which are already known \cite{Grozin:2017css}. We refer the interested reader to the latter reference for the definition of these topologies. The remaining three MI are new and belong to the topologies 4 and 5. One of these three MI is a particular case of an integral calculated in \cite{Grozin:2000jv,Grozin:2003ak}.

 Conveniently we introduce one large set of propagators $\{D_k\}$ to represent the integrals of both two new topologies 4 and 5 and then restrict the exponents of the propagators accordingly. We define
\begin{align} \label{eq:Def_Int_1}
 G(a_1,...,a_{16})=e^{4 \epsilon \gamma_E}\int \! \frac{d^d k_1}{i \pi^{d/2}} \int \! \frac{d^d k_2}{i \pi^{d/2}}  \int \! \frac{d^d k_3}{i \pi^{d/2}}  \int \! \frac{d^d k_4}{i \pi^{d/2}} \prod_{k=1}^{16} D_k^{-a_k} \,,
\end{align}
with the denominators
\begin{align} \label{eq:Def_Int_2}
 & D_1=-2 v \cdot k_1 + 1 \,,		& & D_2=-2 v \cdot k_2 + 1 \,,		& & D_3=-2 v \cdot k_3 + 1 \,,	& \\
 & D_4=-2 v \cdot k_4 + 1 \,,		& & D_5=-k_1^2 \,,			& & D_6=-k_2^2 \,,		&  \nn\\
 & D_7=-k_3^2 \,,			& & D_8=-k_4^2 \,,			& & D_9=-(k_1-k_2)^2 \,,	& \nn\\
 & D_{10}=-(k_1-k_3)^2 \,,		& & D_{11}=-(k_1-k_4)^2 \,, 		& & D_{12}=-(k_2-k_3)^2 \,, 	& \nn\\
 & D_{13} = -(k_2-k_4)^2 \,, 		& & D_{14}=-(k_3-k_4)^2 \,, 		& & D_{15} = -2 v \cdot (k_1-k_2)+1 \,, & \nn\\
 & D_{16}=-2v \cdot(k_1-k_3)+1\,, 	& &					& &				&	\nn
\end{align}
and $v^2=1$. Note that an integral at four loops with our kinematics can have at most 14 linear independent propagators. The restrictions on the (integer) exponents for the two topologies are given by
\begin{align}
 &\text{topology 4: } a_{15},\, a_{16}=0 \; \text{and } a_{10},\, a_{11}, \, a_{12} \ge 0 \, , \hspace{13em} \\
 &\text{topology 5: } a_1,\, a_3=0 \; \text{and } a_4,\, a_5 ,\, a_{10} ,\, a_{11} \ge 0 \,.
\end{align}
The exponents that are not listed are not restricted.

To compute the three new MI we proceed exactly as in \cite{Grozin:2017css}. 
Let us briefly summarize the method here.
By raising the power of the (IR-regulated) Wilson line propagators we choose a basis of integrals that are finite up to
a factorizable overall UV divergence (and trivial divergent factors from bubble-type subdiagrams that we integrate out). To factor out the overall divergence we conveniently work in position space. Using Feynman parameters
we are then left with finite parameter integrals, which we expand to the required order in $\eps$. The individual terms in the $\eps$ expansion are evaluated with the {\tt HyperInt} package~\cite{Panzer:2014caa}. We checked our analytic results for the MI numerically with {\tt FIESTA4} \cite{Smirnov:2015mct}. The analytic results for the MI are presented in the appendix \ref{app:MI}.

\subsection{Renormalization}
Putting all pieces together we get the bare expression for the one-particle-irreducible vertex function $V(\phi)$ and from (\ref{eq:lnW}) the bare expression of $\log W$ up to $\ord{\alpha_s^4}$. 
In order to extract  $\Gamma_\cusp^{(4)}$ using \eq{LogZ} we first need to express the bare $\log W$ in terms of renormalized quantities.
All present $1/\eps^n$ divergences are of UV origin, because we introduced an off-shellness $\delta =-1/2$ in the HQET propagator to regulate the IR divergences. This IR regulator breaks gauge invariance, which becomes evident when computing the off-shell Wilson loop in covariant gauge, i.e. using the gluon propagator
\begin{align} \label{eq:GluonProp}
 \begin{tikzpicture}[scale=1.7, baseline=(current bounding box.center)]	  
	  \draw[Gluon] (0,0)  --  (1.2,0)  {};	  
	  \draw[->] (0.3,-0.2)  --  node[pos=0.5, anchor = north] {$p$} (0.9,-0.2)  {};
\end{tikzpicture}= \frac{-i}{p^2+i0^+} \left( g^{\mu \nu}+\xi \frac{p^\mu p^\nu}{p^2+i0^+}\right) \; .
\end{align}
In fact, not only the finite part, but also some divergent terms of $\log W$ depend on $\xi$.
The latter are related to the interplay of finite and divergent pieces of lower-loop subdiagrams.
Therefore it is crucial to renormalize the gauge parameter $\xi$. We emphasize that the corresponding renormalization of the gauge fixing part of the Lagrangian is necessary even when using Feynman gauge ($\xi=0$) from the start. 
In our convention the gauge parameter renormalizes according to $1-\xi^\textrm{bare}=Z_A (1-\xi)$, where $Z_A$ is the renormalization factor of the gauge field $A_\mu^\textrm{bare}=\sqrt{Z_A}A_\mu$. The required renormalization constants are e.g. given in \cite{Chetyrkin:2017bjc}. 
After expressing $\alpha_s^\textrm{bare}$ and $\xi^\textrm{bare}$ through the respective renormalized quantities the only divergences left in $\log W$ are those associated with the Wilson loop cusp and match \eq{LogZ}. 
As the cusp anomalous dimension is of UV origin and thus insensitive to the off-shellness, it is gauge invariant. Hence, $Z$ must be $\xi$ independent. This serves as a strong check of our calculation. Note that the finite part of $\log W$ does depend on the gauge parameter as well as on the off-shellness.
The determination of the HQET field anomalous dimension $\gamma_h$ from $V(0)$ follows the same lines. Unlike for $\Gamma_\cusp$ in this case, however, a  dependence on $\xi$ persists.

 \subsection{Checks of the calculation}
 \label{sec:checks}
 
 We performed several checks of our calculation. 
Using our computational setup we reproduce the known lower loop results of $\Gamma_\textrm{cusp}(\phi,\alpha_s)$ \cite{Korchemsky:1987wg, Grozin:2015kna} and $\gamma_h(\alpha_s)$ \cite{Broadhurst:1991fz,Melnikov:2000zc,Chetyrkin:2003vi}. In the case of $\gamma_h(\alpha_s)$ our findings at four loops are in agreement with the known analytical and numerical results \cite{Marquard:2018rwx}. Regarding the cusp anomalous dimension, our result for the $n_f C_R C_F^2$ term agrees with \cite{Grozin:2018vdn}. Furthermore, RG consistency and gauge invariance provide stringent tests. Due to the RGE, the higher $1/\eps^n$ poles of $\log W$ in \eq{lnW} are completely determined by the $\beta$-function and lower loop results of the cusp anomalous dimension, see \eq{LogZ}. The same applies to the HQET field anomalous dimension. We use this as a direct check of our four loop computation. As argued before all divergent terms in $\log W$ are gauge invariant. We explicitly verify gauge invariance by computing $\log W$ in covariant gauge up to four loops and observe that the dependence on the gauge parameter $\xi$ drops out in the divergent terms. For $\gamma_h$ this check is not possible, since it is gauge dependent.

\section{Results}
\label{sec:results}
 In this section we collect the our results for the fermionic contributions to the four-loop cusp anomalous dimension $\Gamma_\cusp^{(4)}(\phi)$ in the small angle expansion to $\mathcal{O}(\phi^4)$ or higher. We also give the corresponding contributions to the HQET heavy quark field anomalous dimension $\gamma_h^{(4)}(\alpha_s)$. 

\subsection{HQET field anomalous dimension}
\label{sec:results_field}

We determine $\gamma_h^{(4)}$ from our calculation of the vertex function at zero cusp angle, $V(0)$. According to \eqs{Zh2}{gammadef} we obtain
\begin{align}
&\CSff{\gamma_h}{RA} = \left( -\frac{35 \zeta_3}{24}+\frac{\pi^4}{120}-\frac{4157}{31104} \right)
+\xi \left(-\frac{\zeta_3}{24}+\frac{269}{7776}\right)\,,
\label{eq:ffRA}\\
&\CSf{\gamma_h}{RFA} = \left(-\frac{15 \zeta_5}{8}-\frac{105 \zeta_3}{32}+\frac{23 \pi^4}{960}+\frac{36503}{13824} \right)
+\xi \left(-\frac{11 \zeta_3}{32}-\frac{\pi^4}{960}+\frac{767}{1536}\right)\,,
\label{eq:fRFA}\\
&\CSf{\gamma_h}{RAA} = \left( -\frac{2299 \zeta_5}{1152}-\frac{3 \zeta_3^2}{8}+\frac{37 \pi^2 \zeta_3}{864}+\frac{1751 \zeta_3}{256}-\frac{1529 \pi^4}{69120}-\frac{\pi^2}{48}+\frac{690965}{497664} \right)
\nonumber\\
&\hphantom{\CSf{\gamma_h}{RAA}={}}{}
+ \xi \left(-\frac{\zeta_5}{576}+\frac{\pi^2 \zeta_3}{432}+\frac{65 \zeta_3}{192}-\frac{7 \pi^4}{23040}+\frac{49729}{497664}\right)
\nonumber\\
&\hphantom{\CSf{\gamma_h}{RAA}={}}{}
+ \xi^2 \left(-\frac{7 \zeta_3}{768}+\frac{\pi^4}{46080}-\frac{109}{9216}\right)\,,
\label{eq:fRAA}
\end{align}
where we used the notation of \eq{Color4L}. 
The result for $\CSff{\gamma_h}{RA}$ agrees with the analytic expression obtained in~\cite{Marquard:2018rwx}. The other two results for 
$\CSf{\gamma_h}{RFA}$ and $\CSf{\gamma_h}{RAA}$ agree with the numerical results of~\cite{Marquard:2018rwx}.
Note that the remaining terms can also be found in that reference.

\subsection{Cusp anomalous dimension}
\label{sec:results_cusp}

Regarding $\Gamma_\cusp^{(4)}(\phi)$, the results for the `Abelian' color structures $(n_f T_F)^3C_R$ \cite{Grozin:2004yc}, $(n_f T_F)^2C_R C_F$ \cite{Grozin:2015kna,Grozin:2016ydd} and $(n_f T_F)C_R C_F^2$ \cite{Grozin:2018vdn} are known with full angle dependence and given in \eq{Cusp4LKnown} in the appendix. 
The $\phi^2$ and $\phi^4$ terms of the $n_f d_R d_A/N_R$ contribution were computed in \cite{Grozin:2017css}. Here we extend its small angle expansion to include the $\phi^6$ term, see also \cite{Bruser:2018aud}:
\begin{align}
 \CSf{\Gamma_\text{cusp}}{dRF} = & \phi^2 \left(-\frac{4 \pi ^2 \zeta_3}{9}+\frac{5 \pi ^2}{54}+\frac{5 \pi ^4}{108} \right) \nn\\
 & + \phi^4 \left( \frac{71 \zeta_3}{225}-\frac{16 \pi ^2 \zeta_3}{675}-\frac{4 \zeta_5}{9}-\frac{23}{900}-\frac{157 \pi ^2}{8100}+\frac{49 \pi ^4}{8100} \right) \label{eq:resultfdRF}\\
 &+ \phi^6 \left( \frac{983 \zeta_3}{33075}-\frac{32 \pi ^2 \zeta_3}{11025}-\frac{64 \zeta_5}{1323}+\frac{797}{264600}-\frac{1333 \pi ^2}{595350}+\frac{421 \pi ^4}{595350} \right)+ \mathcal{O}(\phi^8) \, . \nn
 \end{align}
Using the notation of \eq{Color4L} we find for the remaining $n_f$ dependent color structures in the small angle expansion 
 \begin{align}
  \CSff{\Gamma_\text{cusp}}{RA} = &  \phi^2 \left(-\frac{35 \zeta _3}{81}+\frac{7 \pi ^4}{3240}+\frac{19 \pi ^2}{1458}-\frac{1835}{15552}\right)  
	 +\phi^4 \left(-\frac{7 \zeta _3}{243}+\frac{7 \pi^4}{48600}+\frac{19 \pi ^2}{21870}-\frac{5201}{699840}\right)  \nn \\
	& +\phi^6 \left(-\frac{2 \zeta _3}{729}+\frac{\pi ^4}{72900}+\frac{19 \pi ^2}{229635}-\frac{25397}{36741600}\right)  + \mathcal{O}(\phi^8) \;, \label{eq:SMA_ffRA}\\
	\CSf{\Gamma_\text{cusp}}{RFA} = &	 \phi^2 \left(\frac{\pi ^2 \zeta _3}{9}-\frac{85 \zeta _3}{54}-\frac{5 \zeta _5}{12}+\frac{11 \pi ^4}{2160}-\frac{55 \pi ^2}{432}+\frac{25943}{15552}\right) \label{eq:SMA_fRFA}\\
    & + \phi^4 \left(\frac{\pi ^2 \zeta _3}{135}-\frac{41 \zeta _3}{405}-\frac{\zeta _5}{36}+\frac{11 \pi ^4}{32400}-\frac{11 \pi ^2}{1296}+\frac{24953}{233280}\right) + \mathcal{O}(\phi^6) \;, \nn\\
\CSf{\Gamma_\text{cusp}}{RAA} = & \phi^2 \left(-\frac{7}{54} \pi ^2 \zeta _3+\frac{3611 \zeta _3}{1296}-\frac{55 \zeta _5}{72}+\frac{11 \pi ^4}{648}-\frac{923 \pi ^2}{2916}+\frac{48161}{31104}\right) \label{eq:SMA_fRAA} \\
   &+ \phi^4 \left(-\frac{13 \pi ^2 \zeta _3}{1350}+\frac{149327 \zeta _3}{486000}-\frac{47 \zeta _5}{1080}+\frac{293 \pi ^4}{486000}-\frac{35837 \pi ^2}{1749600}+\frac{112207}{34992000}\right) + \mathcal{O}(\phi^6) \;. \nn
 \end{align}

\section{Conjecture on full angle dependence}
\label{sec:conjecture}

Before we compare its predictions to our results of the previous section, let us briefly review the conjecture we want to test.

\subsection{Conjecture}
\label{sec:explainconjecture}

In \cite{Grozin:2014hna,Grozin:2015kna} the authors observed an intriguing pattern in the result of the cusp anomalous dimension at the first three loop orders. To see this pattern we expand $\Gamma_\cusp(\phi)$ in an effective coupling $\lambda= \pi K(\alpha_s)/C_R$ as
\begin{align} \label{eq:conj}
\Gamma_\text{cusp}\big[\phi,\alpha_s(\lambda)\big] = \sum_{k \ge 1} \left( \frac{\lambda}{\pi} \right)^k 
\Omega^{(k)}(\phi)
\, .
\end{align}
In the light-like limit the cusp anomalous dimension equals the lowest order ($k=1$) term in \eq{conj} by construction. All higher order $\lambda$ terms vanish in that limit.
Starting at four loops $\lambda$ depends on the $SU(N_c)$ representation $R$ due to the appearance of the quartic Casimir operators.
In \cite{Grozin:2014hna,Grozin:2015kna} it was found that the expansion coefficients $\Omega^{(k)}(\phi)$ for $k\le 3$ are independent of the matter content of the theory, i.e. the number of scalars ($n_s$) and fermions ($n_f$), see \eqss{omega1}{omega3}.
In particular, they are equal in QCD, pure Yang-Mills, and $\mathcal{N}=4$ sYM theory.  The parameters $n_f$ and $n_s$ enter \eq{conj} only through the light-like cusp anomalous dimension $K(\alpha_s)$, i.e. through $\lambda$. Based on this observation, the authors of \cite{Grozin:2014hna,Grozin:2015kna} conjectured that the coefficients $\Omega^{(k)}(\phi)$ in \eq{conj} are matter-independent to all orders in the $\lambda$ expansion.

This conjecture is particularly interesting because of its predictive power. It allows for predictions on the matter-dependent terms in the loop expansion of $\Gamma_\cusp$ based on lower-loop results. This can be understood by re-expanding \eq{conj} in $\alpha_s$:
\begin{align}
\label{eq:OmegaAlpha}
\Gamma_\cusp(\phi,\alpha_s) =& \; \frac{\alpha_s}{\pi} \Omega^{(1)}(\phi) + \left(\frac{\alpha_s}{\pi} \right)^2 \left[ \Omega^{(2)}(\phi)+ \frac{1}{C_R} K^{(2)}\Omega^{(1)}(\phi) \right] \\
&\; +\left(\frac{\alpha_s}{\pi} \right)^3 \left[ \Omega^{(3)}(\phi)+ \frac{1}{C_R} \left( K^{(3)}\Omega^{(1)}(\phi) + 2 K^{(2)}\Omega^{(2)}(\phi) \right) \right] \nn \\
&\; +\left(\frac{\alpha_s}{\pi} \right)^4 \bigg[ \Omega^{(4)}(\phi)+ \frac{1}{C_R} \left( K^{(4)}\Omega^{(1)}(\phi) + 2 K^{(3)}\Omega^{(2)}(\phi) + 3 K^{(2)}\Omega^{(3)}(\phi)  \right) \nn \\
& \phantom{+\left(\frac{\alpha_s}{\pi} \right)^4} +\frac{1}{C_R^2} \left( K^{(2)} \right)^2 \Omega^{(2)}(\phi)  \bigg] + \mathcal{O}\left( \alpha_s^5\right)  \,, \nn
\end{align}
where we have already inserted the explicit one-loop expression for the light-like cusp anomalous dimension $K^\one=C_R$.
To predict for instance the $n_f$ piece of $\Gamma_\cusp$ at two loops it is sufficient to know $\Omega^{(1)}=C_R(\phi \tan \phi-1)$ and the $n_f$ term of the two-loop light-like cusp anomalous dimension $\CSf{K}{R}=-5 n_f T_F C_R /9$:
\begin{align} \label{eq:conj_fR}
\CSf{\Gamma_\cusp}{R} = n_f T_F \Omega^{(1)} \CSf{K}{R} = -n_f T_F C_R \frac{5}{9}(\phi \tan \phi-1)  \,.
\end{align}
In general, 
at $L$ loops the angle dependence of the $n_f$ contribution to $\Gamma_\cusp$ is completely determined 
by the lower loop coefficients $\Omega^{(k)}(\phi)$ with $k \le L-1$.
In addition some $L$-loop input, e.g. from the asymptotic behavior of $\Gamma_\cusp^f$ in one of the limits, light-like, small angle or anti-parallel lines, is required to fix the constant $K^{(L)}$.

Note that the conjecture does not make any statement on the purely gluonic contributions to the cusp anomalous dimension.
In the following we systematically address the question, for which of the fermionic four-loop terms the conjecture can be successfully validated,  and if some of the yet unknown contributions to $\Gamma_\cusp^{(4)}(\phi)$ can be predicted reliably.
The known $C_A$-independent (`Abelian') fermionic terms with full angle-dependence are easily confirmed to exactly agree with the conjectured results in \eq{Cusp4LKnown}.
For the quartic Casimir color structure the check was performed in \cite{Grozin:2017css} using the terms in the small angle expansion up to $\ord{\phi^4}$.
Here we extend it to the remaining four-loop $n_f$ contributions.

\subsection{Test of the conjecture at small angles}

In \app{conj_res} we give the expressions for the $n_f$ terms in $\Gamma^{(4)}_\cusp(\phi)$ as predicted by the conjecture.
They are obtained by inserting the known lower order $\Omega^{(k)}(\phi)$ results~\cite{Grozin:2014hna} in \eq{OmegaAlpha} and identifying the $n_f$-dependence from the $K^{(i)}$ with $i=2,3,4$. 
Except for the four-loop terms $K^{fRFA}$ and $K^{fRAA}$ associated with the color structures $n_f T_F C_R C_F C_A$ and $n_f T_F C_R C_A^2$, respectively [cf. \eq{Color123L}], for which only numerical results are available in the literature~\cite{Moch:2017uml}, all relevant terms of the $K^{(i)}$ are known analytically, cf.\ \tab{status} and \eqs{K123L}{K4L}.
Now we test the conjectured predictions for the (`non-Abelian') $n_f$ contributions, where the full angle dependence is unkown yet, using our results at small angles.

\subsubsection*{Color structures $n_f T_F C_R C_A^2$ and $n_f d_R d_F/N_R$}

The contribution with the quartic Casimir factor $n_f d_R d_F/N_R$ was studied already in \cite{Grozin:2017css}. 
It was found that the conjectured result for that color structure is incorrect. 
Nevertheless, we repeat here the numerical comparison between conjectured and calculated result in the $\phi$ expansion including the new $\phi^6$:
\begin{align} 
\CSf{\Gamma}{dRF} = & 0.150721 \,\phi^2  + 0.00965191 \,\phi^4 + 0.000925974 \,\phi ^6 + \mathcal{O}(\phi^8) \, , 
\label{eq:fdRFnum}
\\
\CSf{\Gamma_\text{conj.}}{dRF}= & 0.161321 \,\phi^2  + 0.0107548\,\phi^4  + 0.00102426\,\phi^6 + \mathcal{O}(\phi^8) \, .
\label{eq:fdRFconjnum}
\end{align}
The first equation is the numerical version of \eq{resultfdRF}.
The second equation represents the conjectured result, where we have used the recent analytic result for $K^{fdRF}$~\cite{Lee:2019zop,Henn:2019rmi} as the four-loop input.
Instead, in \cite{Grozin:2017css} the overall normalization constant was determined from the anti-parallel lines limit and the quartic Casimir $n_f$ term in the analytic result for the static potential~\cite{Lee:2016cgz}. In that case the conjectured expression is numerically even closer to the correct result in \eq{fdRFnum} than is \eq{fdRFconjnum}.
Regardless of the overall factor the conjectured $\phi$ dependence disagrees (at small angles) with the correct result on the analytical level.

Similarly for the color structure $n_f T_F C_R C_A^2$ the conjecture can be disproved.
Using the $\phi^2$ term of our expanded result in \eq{SMA_fRAA} we can determine the light-like cusp anomalous dimension term $\CSf{K}{RAA}$ in the conjectured all-angles result, \eq{conj_fRAA}, analytically.
The numerical value $\CSf{K}{RAA} = -3.4375$ is quite close to the known numerical result $\CSf{K}{RAA}=-3.4426 \pm0.0016$ of \cite{Moch:2017uml}.
With $\CSf{K}{RAA}$ fixed we obtain an analytical prediction for the $\phi^4$ term of $\CSf{\Gamma}{RAA}$. 
This prediction however contradicts  \eq{SMA_fRAA} despite being numerically close:
\begin{align}
\CSf{\Gamma}{RAA}= & 1.09716 \,\phi^2  + 0.069745 \,\phi^4  + \mathcal{O}(\phi^6) \, , \\
\CSf{\Gamma_\text{conj.}}{RAA}= & 1.09716 \,\phi^2  + 0.069845 \,\phi^4  + \mathcal{O}(\phi^6) \, .
\end{align}

\subsubsection*{Color structures $(n_f T_F)^2 C_R C_A$ and $n_f T_F C_R C_F C_A$}

For the $(n_f T_F)^2 C_R C_A$ structure the light-like cusp anomalous dimension $K^{ffRA}$ is known analytically \cite{Henn:2016men,Davies:2016jie}, cf. equation \eq{K4L}. In the small angle limit we have computed the expansion to $\mathcal{O}(\phi^6)$, see \eq{SMA_ffRA}. This allows for three independent test of the conjectured expression in \eq{conj_ffRA}. We find perfect agreement.

The $n_f T_F C_R C_F C_A$ structure is more complex and there is less analytical
data available.
We have obtained the small angle expansion to $\mathcal{O}(\phi^4)$ in \eq{SMA_fRFA} and the corresponding light-like limit is only known numerically~\cite{Moch:2017uml}.
These results are still sufficient to allow for one analytical and one numerical check in order to validate the conjecture, i.e. \eq{conj_fRFA}.
Using the $\phi^2$ term of our result in \eq{SMA_fRFA} as input we find for the (conjectured) light-like cusp anomalous dimension
\begin{align} \label{eq:Kconj_fRFA}
\CSf{K_\text{conj.}}{RFA} = -\frac{1}{6} \pi ^2 \zeta _3+\frac{29 \zeta _3}{9}+\frac{5 \zeta _5}{4}-\frac{11 \pi ^4}{720}+\frac{55 \pi ^2}{288}-\frac{17033}{5184} = 0.3031\, .
\end{align}
This is in perfect agreement with the known numerical value $K^{fRFA}=0.3027 \pm 0.0016$ \cite{Moch:2017uml}. In addition, with  \eq{Kconj_fRFA} the analytic $\phi^4$ terms of the conjectured and the computed result match exactly.

\subsection{Summary of results and discussion}
\label{sec:discussion}

For the two color structures $n_f T_F C_R C_A^2$ and $n_f d_R d_F/N_R$ the conjecture does not hold. 
We also checked that a redefinition of the quartic Casimir such that a rational fraction of the $n_f T_F C_R C_A^2$ coefficient is shifted to the $n_f d_R d_F/N_R$ coefficient does not change the conclusion.%
\footnote{This includes the particular linear combination obtained in the planar limit ($N_c \to \infty$) of the linear $n_f$ contribution to $\Gamma_\cusp(\phi)$.}
The $n_f T_F C_R C_A^2$ and $n_f d_R d_F/N_R$ terms have in common that, unlike the other fermionic terms, they receive contributions from the diagrams with a one-loop fermion box subdiagram.
The latter first appear at four loops, where they represent the most complicated class of Feynman diagrams.
It is conceivable that only these particular diagrams are responsible for the disagreement with the conjectured results.
This would explain why, at least according to our tests, all fermionic four-loop contributions except for the $n_f T_F C_R C_A^2$ and $n_f d_R d_F/N_R$ terms do agree with the conjecture.

Most interestingly, we have shown that the conjectured 
$(n_f T_F)^2 C_R C_A$ and $n_f T_F C_R C_F C_A$ contributions to $\Gamma_\cusp(\phi)$, given in \eqs{conj_ffRA}{conj_fRFA}, respectively, exactly reproduce both the small $\phi$ expansion as well as the light-like limit ($\phi \to i \infty$).
We think that these exact predictions of complicated analytical results containing transcendental numbers up to weight five strongly supports the conjecture for both of these color structures. While we were able to perform one such analytical test of the $n_f T_F C_R C_F C_A$ term, the $(n_f T_F)^2 C_R C_A$ term even passes three independent analytical tests of that kind. On the other hand, the light-like $n_f T_F C_R C_F C_A$ prediction is in addition checked numerically at the per-mil level.
We remark that unlike for the `Abelian' color structures (without $C_A$), the $\phi$-dependence of these terms is not just given by the one-loop coefficient $\Omega^{(1)}$, but also involves the more complicated coefficient $\Omega^{(2)}$ of the $\lambda$ expansion in \eq{conj}.

Our conjectured result for the $n_f T_F C_R C_F C_A$ contribution includes the important special case of the light-like cusp anomalous dimension $\CSf{K}{RFA}$, for which we provide a novel analytic result in \eq{Kconj_fRFA}.
Based on the evidence found we assume in the following that \eq{Kconj_fRFA} is the correct exact result.
As shown in \cite{Henn:2019rmi}, we are thus in the position to determine also the last missing fermionic contribution $\CSf{K}{RAA}$ analytically by combining the other linear $n_f$ pieces and the known planar $n_f$ term~\cite{Henn:2016men,Moch:2017uml}:
\begin{align}
\CSf{K}{RAA}_\text{conj.} &= 2 K^{(4)}_{\mathrm{planar},n_f}
- \frac{\CSf{K}{RFA}_\text{conj.}}{2}
- \frac{\CSf{K}{RFF}}{4} 
- \frac{\CSf{K}{dRF}}{24} 
\label{eq:KfRAA}
\\
&= -\frac{361 \zeta_3}{54}+\frac{7 \pi ^2
	\zeta_3}{36}+\frac{131 \zeta_5}{72} -\frac{24137}{10368}
+\frac{635 \pi ^2}{1944}-\frac{11 \pi ^4}{2160} 
 \label{eq:Kconj_fRAA} \\
&= - 3.44271 \notag
\,.
\end{align}
To obtain \eq{KfRAA} we have expanded the associated color factors to leading order in $1/N_c$ in order to match the prefactor of $K^{(4)}_{\mathrm{planar},n_f}$.
Again we find perfect agreement of our conjectured result in \eq{Kconj_fRAA} with the numerical result 
$\CSf{K}{RAA} = -3.4426 \pm 0.0016$ of \cite{Moch:2017uml}.

\section{Anti-parallel lines limit}
\label{sec:APL}

In the anti-parallel lines limit the cusp anomalous dimension is closely related to the static quark-antiquark  potential, which was first observed at one loop in \cite{Kilian:1993nk}. 
The expansion around $\delta = \pi - \phi \ll 1$ takes the form
\begin{align}
 \Gamma_\text{cusp}(\pi-\delta,\alpha_s)=- C_R \alpha_s \frac{V_\text{cusp}(\alpha_s)}{\delta}+ \mathcal{O}\left( \alpha_s^4 \frac{\log \delta}{\delta}\right) \;,
\end{align}
where the $\log \delta$ term at four loops is only present in the $C_R C_A^3$ color structure \cite{Bruser:2018aud} and the coefficient $V_\cusp$ is $\delta$-independent. The relation to the static quark-antiquark potential can be understood by interpreting $\delta$ on the right hand side of the equation above as the distance between the static quarks. Indeed, there exits a conformal transformation for $\delta \ll 1$ that maps the cusp configuration to two anti-parallel Wilson lines separated by the distance $\delta$, see e.g.~\cite{Correa:2012nk,Grozin:2015kna}.
In momentum space the static quark-antiquark potential is given by \cite{Anzai:2009tm,Smirnov:2009fh,Lee:2016cgz}
\begin{align}
 V(\vec{q})= - C_R \frac{4 \pi \alpha_s(|\vec{q}|)}{ \vec{q}^{\,2}} V_{Q \bar{Q}}\big(\alpha_s(|\vec{q}|)\big)\,,
\end{align}
where we set the renormalization scale to $\mu= |\vec{q}|$ in order to avoid logarithms of the form $\log^n(\mu^2/\vec{q}^{\,2})$. It is however straightforward to restore the full dependence on the renormalization scale \cite{Smirnov:2008pn}. After Fourier transformation to position space and for equal renormalization scales in the cusp anomalous dimension and the position-space static potential, one directly finds $V_\text{cusp}(\alpha_s)= V_{Q\bar{Q}}(\alpha_s)$ in a conformal theory like $\mathcal{N}=4$ sYM, for details we refer to \cite{Grozin:2015kna}. 

In QCD, however, conformal invariance is broken by an anomaly, which becomes manifest in the running of the strong coupling $\alpha_s$. The relation between $V_\cusp$ and $V_{Q\bar{Q}}$ must therefore be supplemented by terms proportional to the QCD $\beta$-function, $\beta(\alpha_s) = \ord{\alpha_s}$, see e.g. \cite{Braun:2003rp},
\begin{align} \label{eq:CuspQQPot}
 V_\text{cusp}(\alpha_s)-V_{Q\bar{Q}}(\alpha_s)=\beta(\alpha_s) C(\alpha_s)\;, \qquad C(\alpha_s)=\sum_{k \ge 1} \left(\frac{\alpha_s}{\pi} \right)^k C^{(k)} \; .
\end{align}
With the known three-loop cusp anomalous dimension and the two loop static potential \cite{Peter:1996ig,Peter:1997me,Schroder:1998vy} we have $C^{(1)}=(47 C_A-28 n_f T_F)/27$ \cite{Grozin:2015kna}. Note the absence of transcendental terms in this expression. 

Since also the three-loop result for the static potential is available analytically \cite{Lee:2016cgz}, we can use the conjecture to extract information on $C^{(2)}$. Given that the conjecture does not hold for all color structures at four loops, we decompose the cusp anomalous dimension in two terms. The first term is predicted by the conjecture, while the second term accommodates the correction to the conjecture for the $n_f T_F C_R C_A^2$ and $n_f d_R d_F/N_R$ contributions.
Assumig the correctness of the conjectured results for all other four-loop color structures we thus write in the anti-parallel lines limit%
\footnote{Note the (traditional) factor of $1/C_R$ in the definition of the quartic Casimir color structure associated with the $\CSf{V}{dRF}$ potential coefficient. We adopt this convention also in \eqs{VfdRF_1}{VfdRF}.}
\begin{align} \label{eq:Vcusp_corr}
 V_\cusp(\alpha_s) \big|_\mathrm{ferm.}=V_\textrm{conj.}(\alpha_s)-\left(\frac{\alpha_s}{\pi} \right)^3 n_f \left[ \frac{d_R d_F}{N_R C_R} \CSf{V_\textrm{corr.}}{dRF} + T_F C_A^2 \CSf{V_\textrm{corr.}}{RAA}  \right]+ \mathcal{O}(\alpha_s^4)\;.
\end{align} 
According to \eq{CuspQQPot} this information from the fermionic contributions is sufficient to fix
\begin{align}
C^{(2)}={}&(n_f T_F)^2 \left(\frac{134}{243}-\frac{2 \zeta_3}{9}\right)
+ n_f T_F C_A \left(-\frac{5 \zeta _3}{4}-\frac{\pi ^4}{24}+\frac{79}{3888}\right) \\
&+   n_f T_F  C_F\left(\frac{19 \zeta _3}{6}+\frac{\pi ^4}{60}-\frac{1711}{288}\right) 
+ C_A^2 \bigg( 3 \CSf{K}{RAA} -3 \CSf{V_\textrm{corr.}}{RAA}
 -\frac{171 \zeta _3^2}{128} -\frac{211 \pi ^6}{17920} \nn\\
& \qquad -\frac{1091 \zeta _5}{128} -\frac{55 \pi ^2 \zeta_3}{192} 
 +\frac{203 \pi ^4}{1152} +\frac{81 \zeta _3}{4} 
 -\frac{11821 \pi ^2}{5184} +\frac{238315}{31104} 
 +\frac{9}{4} \zeta_{-5,-1} \nn\\
 &\qquad+\frac{21}{32} \pi ^2 \zeta _3 \log (2)
 +\frac{3}{2} \pi ^2 \textrm{Li}_4\left(\frac{1}{2}\right)  +\frac{1}{16} \pi ^2 \log^4(2) 
 -\frac{3}{64} \pi ^4 \log ^2(2) \nn\\
 &\qquad+\frac{5}{192} \pi ^4 \log (2) +\frac{1}{16} \pi ^2 \log (2) \bigg) \;. \nn
\end{align}
The $(n_f T_F)^2$ and $n_f T_F C_F$ terms agree with \cite{Grozin:2016ydd} and the absence of the $C_F^2$ term has also been shown in \cite{Grozin:2018vdn}.
 The transcendental constants $\log(2)$, $\textrm{Li}_4(1/2)$ and $\zeta_{-5,-1}$ come from the $\alpha_s^3 n_f T_F C_A^2$ term in the static potential. 
 Also the quartic Casimir $\alpha_s^3 (n_f d_R d_F)/(N_R C_R)$ term in the static potential contains $\log(2)$ \cite{Lee:2016cgz}, cf.~\eq{VfdRF} below. This is a rather interesting observation, because only the two color structures that disagree with the conjecture contain transcendental constants other than single zeta values.

We can also quantify the corrections in the anti-parallel lines limit for the quartic Casimir $n_f d_R d_F/N_R$ color structure.
The anomaly term $\beta(\alpha_s) C(\alpha_s)$ is independent of the quartic Casimir $n_f d_R d_F/N_R$ at $\mathcal{O}(\alpha_s^3)$, thus we have from \eqs{CuspQQPot}{Vcusp_corr}
\begin{align}
 \CSf{V_\textrm{cusp}}{dRF} = \CSf{K}{dRF}-\CSf{V_\textrm{corr.}}{dRF} = \CSf{V_{Q\tilde{Q}}}{dRF}\,. \label{eq:VfdRF_1}
\end{align}
With the known value for the light-like cusp anomalous dimension $\CSf{K}{dRF}=-0.484$ \cite{Lee:2019zop,Henn:2019rmi} and the three loop static potential \cite{Lee:2016cgz}
\begin{align}
\CSf{V_{Q\tilde{Q}}}{dRF}={}& \frac{5 \pi ^6}{192}-\frac{61 \pi ^2 \zeta _3}{24}-\frac{23 \pi ^4}{48}+\frac{79 \pi ^2}{72} 
+  \log (2)\left(\frac{21}{4} \pi ^2 \zeta _3-\frac{1}{4} \pi ^4 \log(2)+\frac{1}{12} \pi
   ^4+\frac{1}{2} \pi ^2 \right) \nn \\
={}& -0.444 \,. \label{eq:VfdRF}
\end{align}
we see that in the anti-parallel lines limit the correction to the conjecture amounts to $\CSf{V_\textrm{corr.}}{dRF} \approx -10 \%$ for the quartic Casimir $n_f$ contribution.

\section{Conclusion}
\label{sec:conclusion}

In this work we computed the small angle ($\phi$) expansion of fermionic contributions to the QCD cusp anomalous dimension at four loops.
Our results are given in \sec{results}. They include terms up to $\ord{\phi^4}$ and for some of the color structures even up to $\ord{\phi^6}$.
From our calculation for zero cusp angle, i.e. at  $\ord{\phi^0}$, we also obtain new analytic results for the HQET field anomalous dimension in generalized covariant ($\xi$) gauge.

We then used our small angle results for $\Gamma_\cusp(\phi)$ to verify the conjecture of \cite{Grozin:2014hna,Grozin:2015kna}. This conjecture allows to predict the full angle dependence of the fermionic part of $\Gamma_\cusp(\phi)$ from lower-loop results.
Comparing the predicted four-loop expressions, given in \app{conj_res}, to our calculated analytic results at small angle we found strong evidence that the conjectured all-angles expressions are correct for all fermionic contributions except for the $n_f T_F C_R C_A^2$ and the $n_f d_R d_F/N_R$ terms.
The reason for these exceptions might be connected to four-loop HQET Wilson-line diagrams with a fermion box subdiagram, which exclusively contribute to the $n_f T_F C_R C_A^2$ and $n_f d_R d_F/N_R$ pieces. For further discussion see \sec{discussion}.

The conjectured expressions passing our tests include novel results for the $(n_f T_F)^2 C_R C_A$ and $n_f T_F C_R C_F C_A$ contributions to $\Gamma_\cusp(\phi)$ with full angle dependence.
Using the light-like limit of the conjectured $n_f T_F C_R C_F C_A$ term together with available results from the literature we determined in addition novel analytic expressions for the $n_f T_F C_R C_F C_A$ and the $n_f T_F C_R C_A^2$ contributions to the light-like cusp anomalous dimension.
They are in perfect agreement with the known numerical values.
This completes the analytic result for the fermionic part of the four-loop light-like cusp anomalous dimension in QCD~\cite{Henn:2019rmi}.

Finally, we also studied the anti-parallel lines limit of the cusp anomalous dimension. In this limit the cusp anomalous dimension is related to the static quark-antiquark potential plus a conformal anomaly term. The latter is proportional to the QCD $\beta$-function. 
In \sec{APL}, we used our results for $\Gamma_\cusp(\phi)$ 
in order to explore this relationship. We obtained new contributions to the conformal anomaly term
at $\ord{\alpha_s^3}$.

\begin{acknowledgments}
This work was supported in part by a GFK fellowship and by the PRISMA cluster of {excellence} at JGU Mainz.
A.G.'s work has been partially supported by the Russian Ministry of Education and Science. R.B. thanks G. Korchemsky for valuable insghts and suggestions regarding the anti-parallel lines limit, and the Institut de Physique Th\'{e}orique (IPhT) at CEA Saclay for hospitality. R.B.'s work has been partially supported by the Deutsch-Franz\"osische Hochschule under the grant number CDFA-02-18, through a PhD exchange program in theoretical physics, \url{http://dfdk-physik.de}. This research received funding from the European Research Council (ERC) under the European Union's Horizon 2020 research and innovation programme (grant agreement No 725110), {\it Novel
structures in scattering amplitudes}.
The authors gratefully acknowledge support by the HPC group at JGU Mainz.
\end{acknowledgments}

\begin{appendix}
 \section{Lower-loop and conjectured four-loop $\Gamma_\cusp(\phi)$}
 \label{app:conj_res}
 
In this appendix we list the known and conjectured full angle-dependent results for all fermionic color structures of the cusp anomalous dimension at four loops. The results are expressed in terms of the light--like cusp anomalous dimension $K(\alpha_s)$ and seven coefficient functions $A_{1,2,3,4,5}$, and $B_{3,5}$, encoding the angle dependence. For convenience we also present the expansions of the coefficient functions in the small angle and the anti--parallel lines limits. 
We start with the expansion coefficients of $\Gamma_\mathrm{cusp}[\phi,\alpha_s(\lambda)]$ in \eq{conj} as given in \cite{Grozin:2015kna},
\begin{align}
 \Omega^{(1)}(\phi)&= C_R  \tilde{A}_1 \,, \label{eq:omega1}\\
 \Omega^{(2)}(\phi)&= \frac{C_R C_A}{2} \left( \frac{\pi^2 \tilde{A}_1}{6}+\tilde{A}_2 + \tilde{A}_3 \right) \,,\\
 \Omega^{(3)}(\phi)&= \frac{C_R C_A^2}{4} \left(-\tilde{A}_2 +\tilde{A}_4+\tilde{A}_5+\tilde{B}_3+\tilde{B}_5-\frac{\pi^4}{180}\tilde{A}_1+\frac{\pi^2}{3} \left( \tilde{A}_2+\tilde{A}_3\right)\right) \,,
 \label{eq:omega3}
\end{align}
where $\tilde{A}_i =\tilde{A}_i(x)$ and $\tilde{B}_i =\tilde{B}_i(x)$ with $x=e^{i \phi}$. The `Abelian' $n_f$-dependent contributions to $\Gamma_\cusp(\phi)$ are known. They have the same functional angle dependence as the one-loop result \cite{Grozin:2004yc,Grozin:2015kna,Grozin:2016ydd,Grozin:2018vdn}:
 \begin{align} \label{eq:Cusp4LKnown}
  \CS{\Gamma_\text{cusp}}{\mathcal{C}}={}& \CS{K}{\mathcal{C}} \tilde{A}_1 \; , \qquad \text{with } \mathcal{C} = f\!f\!f\!R,\, f\! R F,\, f\!R F F \; .
 \end{align}
Here and in the following we use the notation of \eqs{Color123L}{Color4L} for the different color structures. Next we give the results for the other fermionic color structures as predicted by the conjecture:
\begin{align}
 \CSff{\Gamma_\text{conj.}}{RA}={}& \CSff{K}{RA} \tilde{A}_1 + \left(\frac{\left(\CSf{K}{R} \right)^2}{2} + \CSff{K}{R}\right) \left( \frac{\pi^2 \tilde{A}_1}{6}+\tilde{A}_2 + \tilde{A}_3 \right) , \label{eq:conj_ffRA}\\
 \CSf{\Gamma_\text{conj.}}{RFA}={}&\CSf{K}{RFA} \tilde{A}_1+\CSf{K}{RF} \left( \frac{\pi^2 \tilde{A}_1}{6}+\tilde{A}_2 + \tilde{A}_3 \right) ,\label{eq:conj_fRFA} \\
 \CSf{\Gamma_\text{conj.}}{RAA}={}&  \CSf{K}{RAA} \tilde{A}_1 + \CSf{K}{RA} \left( \frac{\pi^2 \tilde{A}_1}{6}+\tilde{A}_2 + \tilde{A}_3 \right) 
  + \CSf{K}{R} \left[- \frac{\pi^4}{240} \tilde{A}_1 \right. \label{eq:conj_fRAA} + \frac{\pi^2}{4} \left( \tilde{A}_2 + \tilde{A}_3 \right) \\
  &+ \frac{3}{4} \left( -\tilde{A}_2+\tilde{A}_4+\tilde{A}_5+\tilde{B}_3+\tilde{B}_5\right)
  \left. +\CS{K}{RA} \left( \frac{\pi^2 \tilde{A}_1}{6}+\tilde{A}_2 + \tilde{A}_3 \right) \right], \nn\\
  \CSf{\Gamma_\text{conj.}}{dRF}={}& \CSf{K}{dRF} \tilde{A}_1\; . 
  \label{eq:conj_fdRF}
\end{align}
Analytical expressions for the light-like cusp anomalous dimension are available  up to three loops \cite{Korchemsky:1987wg,Moch:2004pa}:
\begin{align} 
K={}&C_R \frac{\alpha_s}{\pi} \nn
+\left(\frac{\alpha_s}{\pi}\right)^2 \left[ -\frac{5}{9}  n_f T_F C_R + C_A C_R \left(\frac{67}{36}-\frac{\pi ^2}{12}\right) \right]
+ \left(\frac{\alpha_s}{\pi}\right)^3  \left[ -\frac{1}{27} (n_f T_F)^2 C_R \right. \\
&+ n_f T_F C_R C_F  \left(\zeta_3-\frac{55}{48}\right)
+  n_f T_F C_R C_A \left(-\frac{7 \zeta_3}{6}-\frac{209}{216}+\frac{5 \pi ^2}{54}\right) \label{eq:K123L} \\
&+ \left. C_R C_A^2\left(\frac{11 \zeta_3}{24}+\frac{245}{96}-\frac{67 \pi ^2}{216}+\frac{11 \pi ^4}{720}\right) \right] 
+\left(\frac{\alpha_s}{\pi}\right)^4 K^{(4)}
+ \mathcal{O}(\alpha_s^5)
\,. \nn
\end{align}
Depending on the color factor the contributions at  four loops are either known analytically or numerically  \cite{Beneke:1995pq,Henn:2016men,Davies:2016jie,Lee:2016ixa,Moch:2017uml,Moch:2018wjh,Lee:2019zop,Henn:2019rmi}:
\begin{align}
 K^{(4)}&={}
 (n_f T_F)^3 C_R \left(-\frac{1}{81}+\frac{2 \zeta _3}{27}\right)
 + (n_f T_F)^2 C_R C_A \left( \frac{35 \zeta _3}{27}-\frac{7 \pi ^4}{1080}-\frac{19 \pi ^2}{972}+\frac{923}{5184} \right)  \label{eq:K4L}\\
 &+ (n_f T_F)^2 C_R C_F \left(-\frac{10 \zeta _3}{9}+\frac{\pi ^4}{180}+\frac{299}{648}\right) 
 + n_f T_F C_R C_F^2 \left( \frac{37 \zeta _3}{24}-\frac{5 \zeta _5}{2}+\frac{143}{288} \right) \nn\\
 & +  n_f T_F C_R C_F C_A \big(0.3027 \pm 0.0016 \big)
 + n_f T_F C_R C_A^2 \big(-3.4426 \pm 0.0016 \big) \nn\\
 &+ n_f \frac{d_R d_F}{N_R} \left( \frac{\pi^2}{6}-\frac{\zeta_3}{3}-\frac{5\zeta_5}{3} \right)
 + \frac{d_R d_A}{N_R} \big(-1.9805 \pm 0.0078 \big)
 + C_R C_A^3 \big(2.38379 \pm 0.00039 \big) . \nn
\end{align}
The coefficient functions are given by \cite{Grozin:2015kna}
\begin{align*}
\tilde{A}_i(x) &= A_i(x)-A_i(x)\;, \qquad \tilde{B}_i(x) = B_i(x)-B_i(x)\;, \tagaligneq \\
{A}_1(x) &=  \tilde{\xi} \,   \frac{1}{2} H_{1}(y)\,,\\
{A}_2(x)  &=  \left[ \frac{\pi^2}{3} + \frac{1}{2} H_{1,1}(y) \right]  + \tilde{\xi} \left[- H_{0,1}(y) -\frac{1}{2} H_{1,1}(y) \right] \,, \\
{A}_{3}(x) &=  \tilde{\xi} \, \left[ -\frac{\pi^2}{6} H_{1}(y) - \frac{1}{4} H_{1,1,1}(y)  \right]  
 +  \tilde{\xi}^2\, \left[ \frac{1}{2} H_{1,0,1}(y) + \frac{1}{4} H_{1,1,1}(y) \right]  \,,\\
{A}_{4}(x) &=
 \left[  -\frac{\pi^2}{6} H_{1,1}(y) -\frac{1}{4} H_{1,1,1,1}(y)  \right]  + \tilde{\xi} \left[  \frac{\pi^2}{3} H_{0,1}(y) +\frac{\pi^2}{6} H_{1,1}(y) + 2 H_{1,1,0,1}(y)   \right. \\ 
 &\quad \left.  
 +\frac{3}{2} H_{0,1,1,1}(y) 
 + \frac{7}{4} H_{1,1,1,1}(y) + 3 \zeta_3 H_{1}(y)  \right]  +  \tilde{\xi}^2 \left[ -2 H_{1,0,0,1}(y)-2 H_{0,1,0,1}(y)
  \right. \\
&\quad \left.  -2 H_{1,1,0,1}(y)- H_{1,0,1,1}(y) - H_{0,1,1,1}(y) -\frac{3}{2} H_{1,1,1,1}(y) \right]\,,\\
{A}_{5}(x) &= \tilde{\xi} \left[  \frac{\pi^4}{12} H_{1}(y) + \frac{\pi^2}{4} H_{1,1,1}(y) +\frac{5}{8}  H_{1, 1, 1, 1, 1}(y) \right] +  \tilde{\xi}^2\left[ -\frac{\pi^2}{6} H_{1,0,1}(y) -\frac{\pi^2}{3} H_{0,1,1}(y)  \right.  \\
&\quad \left. 
-\frac{\pi^2}{4}H_{1,1,1}(y)  -H_{1,1,1,0,1}(y)  -\frac{3}{4} H_{1,0,1,1,1}(y) -H_{0,1,1,1,1}(y) -\frac{11}{8} H_{1,1,1,1,1}(y)    \right.  \\
&\quad \left. 
 -\frac{3}{2}\zeta_3 H_{1,1}(y)  \right] 
 + \tilde{\xi}^3 \left[ H_{1,1,0,0,1}(y) + H_{1,0,1,0,1}(y) +H_{1,1,1,0,1}(y) + \frac{1}{2} H_{1,1,0,1,1}(y) 
   \right.  \\
&\quad \left. 
 +\frac{1}{2} H_{1,0,1,1,1}(y) +\frac{3}{4} H_{1,1,1,1,1}(y)  \right] \,,
 \\ 
B_{3}(x) &= \left[  - H_{1,0,1}(y) + \frac{1}{2} H_{0,1,1}(y) - \frac{1}{4} H_{1,1,1}(y)\right]   \\
&\quad + \tilde{\xi} \left[ 2 H_{0,0,1}(y) + H_{1,0,1}(y) + H_{0,1,1}(y) + \frac{1}{4} H_{1,1,1}(y) \right]\,,  \\
B_{5}(x) &= 
\frac{x}{1-x^2} \left[-\frac{\pi^4}{60} H_{-1}(x) -\frac{\pi^4}{60} H_{1}(x) - 4H_{-1,0,-1,0,0}(x) + 4 H_{-1,0,1,0,0}(x)   \right.  \\
& \left. \qquad
 - 4 H_{1,0,-1,0,0}(x)   
+ 4 H_{1,0,1,0,0}(x) + 4 H_{-1,0,0,0,0}(x) + 4 H_{1,0,0,0,0}(x) \right.  \\ 
&  \qquad + 2 \zeta_3  H_{-1,0}(x) + 2 \zeta_3 H_{1,0}(x)   \bigg] \,, 
\end{align*} 
 with $\tilde \xi=(1+x^2)/(1-x^2)$, $y=1-x^2$, $x=e^{i \phi}$. The $H_{\vec{a}}(y)$ denote harmonic polylogarithms according to \cite{Remiddi:1999ew,Maitre:2005uu}.

For the small angle expansion of the above coefficient functions around $\phi=0$ we find up to $\ord{\phi^6}$:
\begin{align*}
A_1(\phi)&=1-\frac{1}{3} \phi^2-\frac{1}{45} \phi^4-\frac{2}{945} \phi^6 +\mathcal{O}(\phi^8) \tagaligneq \;,\\
A_2(\phi)&=\frac{\pi^2}{3}-2-\frac{1}{9} \phi^2-\frac{14}{675} \phi^4-\frac{304}{99225} \phi^6 +\mathcal{O}(\phi^8) \;,\\
A_3(\phi)&=\left(1-\frac{\pi ^2}{3}\right)
+\phi ^2\left(\frac{\pi ^2}{9}-\frac{7}{18}\right) 
+\phi ^4 \left(\frac{\pi ^2}{135}-\frac{2}{225}\right) 
+\phi ^6 \left(\frac{38}{99225}+\frac{2 \pi ^2}{2835}\right) +\mathcal{O}(\phi^8)  \;,\\
A_4(\phi)&= \left(6 \zeta_3+\frac{2 \pi ^2}{3}-6 \right)
+\phi ^2 \left(-2 \zeta_3+\frac{91}{54}+\frac{\pi ^2}{27}\right) 
+\phi ^4 \left(-\frac{2 \zeta_3}{15}+\frac{1789}{20250}+\frac{14 \pi ^2}{2025}\right)  \\
&\quad+\phi ^6\left(-\frac{4 \zeta_3}{315}+\frac{250121}{20837250}+\frac{304 \pi ^2}{297675}\right) 
+\mathcal{O}(\phi^8)\;,\\
A_5(\phi)&= \left(-3 \zeta_3+\frac{\pi ^4}{6}-\frac{2 \pi ^2}{3}+2 \right)
+\phi ^2 \left(2 \zeta_3-\frac{65}{54}+\frac{5 \pi ^2}{27}-\frac{\pi ^4}{18}\right) 
+\phi ^4 \left(-\frac{\zeta_3}{5}+\frac{1649}{10125} \right. \\
&\quad \left. +\frac{41 \pi ^2}{2025}-\frac{\pi^4}{270}\right)  
+\phi^6 \left(-\frac{2 \zeta_3}{63}+\frac{6401}{1157625}+\frac{349 \pi^2}{99225}-\frac{\pi^4}{2835}\right) 
+\mathcal{O}(\phi^8) \;,\\
B_3(\phi)&=4 -\frac{5}{54}\phi^2 -\frac{889 }{40500}\phi^4 -\frac{80299 }{20837250}\phi^6 +\mathcal{O}(\phi^8)\;, \\
 B_5(\phi)&=\frac{3 \zeta_3}{2} 
   +\phi^2 \left(\frac{\zeta_3}{3}+\frac{1}{18}\right) 
   +\phi^4 \left(\frac{11 \zeta_3}{225}+\frac{31}{2700}\right)
   +\phi^6 \left(\frac{202 \zeta_3}{33075}+\frac{143}{99225}\right) 
   +\mathcal{O}(\phi^8)\;.
\end{align*} 
In the anti-parallel lines limit $\delta = \pi - \phi \ll 1$ we obtain:
\begin{align*}
  \delta A_1(\pi-\delta)&= -\pi + \mathcal{O}(\delta) \;,\tagaligneq\\
  \delta A_2(\pi-\delta)&= 2 \pi  \log (i \delta )-i \pi ^2+2 \pi  \log (2) + \mathcal{O}(\delta)\;, \\
  \delta A_3(\pi-\delta)&= -2 \pi  \log (i \delta )+i \pi ^2+2 \pi -2 \pi  \log (2)+ \mathcal{O}(\delta)\;, \\
  \delta A_4(\pi-\delta)&= \frac{\pi ^4}{3 \delta }+4 \pi  \log ^2(i \delta )+\left(-4 i \pi ^2+\frac{4 \pi ^3}{3}+8 \pi  \log (2)\right) \log (i \delta ) \\
  &\quad +9 \pi  \zeta_3-\frac{2 i \pi ^4}{3}-2 \pi ^3-8 \pi +4 \pi  \log^2(2)+\left(\frac{4 \pi ^3}{3}-4 i \pi ^2\right) \log (2) + \mathcal{O}(\delta) \;,\\
 \delta A_5(\pi-\delta)&= -\frac{\pi ^4}{3 \delta }-2 \pi  \log ^2(i \delta )+\left(2 \pi +2 i \pi ^2-\frac{4 \pi ^3}{3}-4 \pi  \log (2)\right) \log (i \delta ) -9 \pi  \zeta_3 \\
 &\quad +\frac{2 i \pi ^4}{3}+\pi ^3-i \pi ^2+3 \pi -2 \pi \log ^2(2)+\left(2 \pi +2 i \pi ^2-\frac{4 \pi ^3}{3}\right) \log (2) + \mathcal{O}(\delta) \;,\\
 \delta B_3(\pi-\delta)&=-2 \pi  \log ^2(i \delta )+\left(-4 \pi  \log (2)+2 i \pi ^2\right) \log (i \delta )-2 \pi  \log ^2(2)+2 i \pi ^2 \log (2) + \mathcal{O}(\delta) \;,\\
 \delta B_5(\pi-\delta)&=\frac{\pi ^5}{16} + \mathcal{O}(\delta) \; .
\end{align*}

  \section{Master integrals}
 \label{app:MI}

 For the small angle expansion of the fermionic part of the cusp anomalous dimension 46 master integrals are needed; 43 of which are already known \cite{Grozin:2017css}. Below we list the three new master integrals, which we compute with the method outlined in section \ref{sec:Integrals_Topos}. These integrals are defined by the \eqs{Def_Int_1}{Def_Int_2}. The first integral is associated with topology 3 and the other two integrals are associated with topology 5 in \fig{integral_topos}.
  \begin{align*}
& G(1,1,1,0,1,1,1,0,0,0,0,0,2,1,0,0)= \tagaligneq
\frac{1}{\epsilon^3} \frac{\pi ^2}{18}
+ \frac{1}{\epsilon^2} \bigg(\frac{2 \pi ^2}{9}-\frac{7 \zeta _3}{3} \bigg) \\ 
& \qquad + \frac{1}{\epsilon} \bigg(-\frac{28 \zeta _3}{3}+\frac{181 \pi ^4}{540}+\frac{2 \pi ^2}{3} \bigg) 
 + \bigg(-\frac{251}{27} \pi ^2 \zeta _3-28 \zeta _3-\frac{250 \zeta _5}{3}+\frac{181 \pi^4}{135}+\frac{16 \pi ^2}{9} \bigg) \\
& \qquad  + \epsilon \bigg( \frac{910 \zeta _3^2}{9}-\frac{1004 \pi ^2 \zeta _3}{27}-\frac{224 \zeta_3}{3}-\frac{1000 \zeta _5}{3}+\frac{1711 \pi ^6}{1260}+\frac{181 \pi^4}{45}+\frac{40 \pi ^2}{9} \bigg) \\
& \qquad + \epsilon^2 \bigg( \frac{3640 \zeta _3^2}{9}-\frac{11617 \pi ^4 \zeta _3}{405}-\frac{1004 \pi ^2 \zeta _3}{9}-\frac{560 \zeta _3}{3}-\frac{4634 \pi ^2 \zeta _5}{15}-1000 \zeta_5 \\
& \qquad \phantom{\epsilon^2 \bigg(+}-\frac{14729 \zeta _7}{6}+\frac{1711 \pi ^6}{315}+\frac{1448 \pi^4}{135}+\frac{32 \pi ^2}{3} \bigg)
+ \mathcal{O}(\epsilon^3)\,,\\
& G(0,1,0,0,0,1,1,0,1,0,0,0,1,2,1,2)= \tagaligneq
\frac{1}{\epsilon^2} \bigg( -\frac{\pi ^2}{12} \bigg) 
+ \frac{1}{\epsilon} \bigg( \frac{9 \zeta _3}{4}+\frac{\pi ^2}{3} \bigg) \\
& \qquad  + \bigg( -31 \zeta _3-\frac{277 \pi ^4}{720}-\pi ^2 \bigg) 
+ \epsilon \bigg( \frac{487 \pi ^2 \zeta _3}{36}+27 \zeta _3-\frac{5 \zeta _5}{4}+\frac{883 \pi^4}{180}+\frac{16 \pi ^2}{3} \bigg) \\
& \qquad + \epsilon^2 \bigg( -\frac{1505 \zeta _3^2}{4}-\frac{1753 \pi ^2 \zeta _3}{9}-408 \zeta _3-2405 \zeta_5-\frac{703 \pi ^6}{1680}-\frac{277 \pi ^4}{60}-\frac{52 \pi ^2}{3} \bigg) \\
& \qquad +\epsilon^3 \bigg( \frac{9085 \zeta _3^2}{3}+\frac{98243 \pi ^4 \zeta _3}{1080}+\frac{487 \pi ^2 \zeta _3}{3}+820 \zeta _3-\frac{23821 \pi ^2 \zeta _5}{60}-15 \zeta _5 \\
& \qquad \phantom{\epsilon^3 \bigg(+} -4184 \zeta _7 +\frac{164417 \pi ^6}{3780}+\frac{2926 \pi ^4}{45}+80 \pi ^2 \bigg) + \mathcal{O}(\epsilon^4)  \,,\\
& G(0,2,0,0,0,0,1,2,1,0,0,0,1,0,2,3)=  \tagaligneq
 \frac{1}{\epsilon} \bigg( \frac{1}{2}-\frac{\pi ^2}{12} \bigg)+ \bigg( 6 \zeta _3-\frac{2 \pi ^2}{3}+3 \bigg) \\
& \qquad + \epsilon \bigg( 48 \zeta _3-\frac{167 \pi ^4}{180}-\frac{5 \pi ^2}{2}+14 \bigg)
 + \epsilon^2 \bigg( \frac{325 \pi ^2 \zeta _3}{9}+\frac{898 \zeta _3}{3}+372 \zeta _5-\frac{334 \pi^4}{45}-\frac{29 \pi ^2}{3}+60 \bigg) \\
& \qquad + \epsilon^3 \bigg( -584 \zeta _3^2+\frac{2600 \pi ^2 \zeta _3}{9}+1412 \zeta _3+2976 \zeta_5-\frac{1733 \pi ^6}{252}-\frac{4109 \pi ^4}{90}-38 \pi ^2+248 \bigg) \\
& \qquad + \epsilon^4 \bigg( -4672 \zeta _3^2+\frac{25403 \pi ^4 \zeta _3}{135}+\frac{5494 \pi ^2 \zeta_3}{3}+\frac{18232 \zeta _3}{3}+\frac{31891 \pi ^2 \zeta _5}{15} \\ 
& \qquad \phantom{\epsilon^4 \bigg(+}  +\frac{102658 \zeta _5}{5}+20052 \zeta _7-\frac{3466 \pi ^6}{63}-\frac{1931 \pi^4}{9}-\frac{452 \pi ^2}{3}+1008 \bigg)
  + \mathcal{O}(\epsilon^5) \,.
\end{align*}

Using IBP relations, the three master integrals can be exchanged for integrals with uniformal transcendental weight:
  \begin{align*}
& G(1, 1, 1, 0, 2, 1, 1, 0, 0, 0, 0, 0, 2, 1, 0, 0)= \frac{1}{\epsilon^3(1-2\epsilon)} \Bigg[ \tagaligneq
 \frac{\pi ^2}{9} - \epsilon \frac{14 \zeta _3}{3}  + \epsilon^2 \frac{181 \pi ^4}{270} \\
 & \qquad \qquad + \epsilon^3 \left(  -\frac{502}{27} \pi ^2 \zeta _3-\frac{500 \zeta _5}{3} \right) 
	  + \epsilon^4 \left( \frac{1820 \zeta _3^2}{9}+\frac{1711 \pi ^6}{630} \right) \\
 & \qquad \qquad +  \epsilon^5 \left(  -\frac{23234}{405} \pi ^4 \zeta _3-\frac{9268 \pi ^2 \zeta _5}{15}-\frac{14729 \zeta _7}{3} \right) + \mathcal{O}(\epsilon^6) \Bigg] \,,\\
& G(0, 1, 0, 0, 0, 0, 1, 2, 2, 0, 0, 0, 2, 0, 1, 2)= \frac{1}{\epsilon^4} \Bigg[ \tagaligneq
-\frac{1}{2}- \epsilon^2 \frac{13 \pi^2}{6} + \epsilon^3 \frac{110 \zeta_3}{3}- \epsilon^4 \frac{63 \pi^4}{10}\\
& \qquad \qquad  + \epsilon^5 \left( \frac{1718 \pi ^2 \zeta _3}{9}+\frac{1502 \zeta _5}{5} \right) 
	  + \epsilon^6 \left(-\frac{22468 \zeta _3^2}{9}-\frac{233 \pi ^6}{15} \right) \\
& \qquad \qquad  + \epsilon^7 \left(\frac{12274 \pi ^4 \zeta _3}{15}+\frac{23366 \pi ^2 \zeta _5}{15}-\frac{74338 \zeta _7}{7} \right) 
	  + \mathcal{O}(\epsilon^8) \Bigg] \,,\\
& G(0, 1, 0, 0, 0, 1, 1, 0, 2, 0, 0, 0, 1, 2, 1, 1)=  \frac{1}{\epsilon^3(1-2\epsilon)} \Bigg[ \tagaligneq
 \frac{\pi^2}{9} - \epsilon \frac{11 \zeta_3}{3}+ \epsilon^2 \frac{383 \pi^4}{540} \\
 & \qquad \qquad + \epsilon^3 \left( -\frac{709}{27} \pi ^2 \zeta _3-\frac{335 \zeta _5}{3} \right) 
	  + \epsilon^4 \left( \frac{4625 \zeta _3^2}{9}+\frac{35437 \pi ^6}{11340} \right) \\
  &  \qquad \qquad + \epsilon^5 \left( -\frac{128977}{810} \pi ^4 \zeta _3-\frac{7693 \pi ^2 \zeta _5}{15}-\frac{18037
   \zeta _7}{6} \right)  
   + \mathcal{O}(\epsilon^6) \Bigg] \,.
\end{align*}

\end{appendix}

\phantomsection
\addcontentsline{toc}{section}{References}
\bibliographystyle{jhep}
\bibliography{Cusp}

\end{document}

%% file: Plots/Dia_fffR.tex
\begin{tikzpicture}[scale=0.8, baseline=(current bounding box.center)]
      \tikzmath{\yPos1 = -1.6;}

      \node at (0,0) {~};
      \node at (0,-2.1) {~};

      \coordinate (WL) at (-2,-2)   {};
      \coordinate (cusp) at (0,0)   {};
      \coordinate (WR) at (2,-2) {};
      
      \coordinate (B1) at (-1.1,\yPos1) {};
      \coordinate (B2) at (-0.7,\yPos1) {};
      \coordinate (B3) at (-0.2,\yPos1) {};
      \coordinate (B4) at (0.2,\yPos1) {};
      \coordinate (B5) at (0.7,\yPos1) {};
      \coordinate (B6) at (1.1,\yPos1) {};
      
      \coordinate (W1) at (-1.6,\yPos1) {};
      \coordinate (W2) at (1.6, \yPos1) {};
      
      \draw[Wilson_1] (WL)  -- (cusp) -- (WR);
      
      \draw[Gluon_1] (W2) -- (B6) ;
      \draw (B6) arc [start angle=0, end angle=360, x radius=0.2, y radius=0.2];
      \draw[Gluon_1] (B5) -- (B4) ;
      \draw (B4) arc [start angle=0, end angle=360, x radius=0.2, y radius=0.2];
      \draw[Gluon_1] (B3) -- (B2) ;
      \draw (B2) arc [start angle=0, end angle=360, x radius=0.2, y radius=0.2];
      \draw[Gluon_1] (B1) -- (W1) ;

\end{tikzpicture}

%% file: Plots/Dia_ffRF.tex
\begin{tikzpicture}[scale=0.8, baseline=(current bounding box.center)]
      \tikzmath{\yPos1 = -1.6;}

      \node at (0,0) {~};
      \node at (0,-2.1) {~};

      \coordinate (WL) at (-2,-2)   {};
      \coordinate (cusp) at (0,0)   {};
      \coordinate (WR) at (2,-2) {};
      
      \coordinate (B1) at (-1.1,\yPos1) {};
      \coordinate (B2) at (-0.7,\yPos1) {};
      \coordinate (B3) at (-0.2,\yPos1) {};
      \coordinate (B4) at (1,\yPos1) {};
     
     \coordinate (B5) at (0.4,-1.3) {};
      \coordinate (B6) at (0.4,-1.9) {};
      
      \coordinate (W1) at (-1.6,\yPos1) {};
      \coordinate (W2) at (1.6, \yPos1) {};
      
      \draw[Wilson_1] (WL)  -- (cusp) -- (WR);
      
      \draw[Gluon_1] (W2) -- (B4) ;
      \draw (B4) arc [start angle=0, end angle=360, x radius=0.6, y radius=0.3];
      \draw[Gluon_1] (B3) -- (B2) ;
      \draw (B2) arc [start angle=0, end angle=360, x radius=0.2, y radius=0.2];
      \draw[Gluon_1] (B1) -- (W1) ;
      
      \draw[Gluon_1] (B5) -- (B6) ;            
\end{tikzpicture}

%% file: Plots/Dia_ffRA.tex
\begin{tikzpicture}[scale=0.8, baseline=(current bounding box.center)]

      \node at (0,0) {~};
      \node at (0,-2.1) {~};

      \coordinate (WL) at (-2,-2)   {};
      \coordinate (cusp) at (0,0)   {};
      \coordinate (WR) at (2,-2) {};
      
      \coordinate (B1) at (-1.1,-1.9) {};
      \coordinate (B2) at (-0.1,-1.9) {};
      \coordinate (B3) at (-0.6,-1.4) {};
      \coordinate (B4) at (0.25,-1.55) {};
      \coordinate (B5) at (0.55,-1.25) {};
      
      \coordinate (W1) at (-1.5,-1.5) {};
      \coordinate (W2) at (-1,-1) {};
      \coordinate (W3) at (0.9, -0.9) {};
      
      \draw[Wilson_1] (WL)  -- (cusp) -- (WR);
      
      \draw[Gluon_1] (B1) -- (W1) ;
      \draw (B1) -- (B2) -- (B3) -- (B1);
      \draw[Gluon_1] (B3) -- (W2) ;
      \draw[Gluon_1] (B4) -- (B2) ;
      \draw[Gluon_1] (W3) -- (B5) ;
      \draw (B5) arc [start angle=45, end angle=405, x radius=0.211, y radius=0.211];
      
\end{tikzpicture}

%% file: Plots/Dia_fRFF.tex
\begin{tikzpicture}[scale=0.8, baseline=(current bounding box.center)]
      \tikzmath{\yPos1 = -1.6;}

      \node at (0,0) {~};
      \node at (0,-2.1) {~};

      \coordinate (WL) at (-2,-2)   {};
      \coordinate (cusp) at (0,0)   {};
      \coordinate (WR) at (2,-2) {};
      
      \coordinate (B1) at (-1,\yPos1) {};
      \coordinate (B2) at (0.4,-1.9) {};
      \coordinate (B3) at (-0.4,-1.9) {};
      \coordinate (B4) at (-0.4,-1.3) {};     
      \coordinate (B5) at (0.4,-1.3) {};
      \coordinate (B6) at (1,\yPos1) {};
      
      \coordinate (W1) at (-1.6,\yPos1) {};
      \coordinate (W2) at (1.6, \yPos1) {};
      
      \draw[Wilson_1] (WL)  -- (cusp) -- (WR);
      
      \draw[Gluon_1] (W2) -- (B6) ;
      \draw (B5) arc [start angle=90, end angle=-90, x radius=0.6, y radius=0.3];
      \draw (B5) -- (B4) ;
      \draw (B2) -- (B3) ;
      \draw (B4) arc [start angle=90, end angle=270, x radius=0.6, y radius=0.3];
      \draw[Gluon_1] (B1) -- (W1) ;
      
      \draw[Gluon_1] (B5) -- (B2) ;
      \draw[Gluon_1] (B4) -- (B3) ;
\end{tikzpicture}

%% file: Plots/Dia_fRFA.tex
\begin{tikzpicture}[scale=0.8, baseline=(current bounding box.center)]

      \node at (0,0) {~};
      \node at (0,-2.1) {~};

      \coordinate (WL) at (-2,-2)   {};
      \coordinate (cusp) at (0,0)   {};
      \coordinate (WR) at (2,-2) {};
      
      \coordinate (B1) at (-0.5,-1.9) {};
      \coordinate (B2) at (+0.5,-1.9) {};
      \coordinate (B3) at (0,-1.4) {};
      \coordinate (B4) at (0,-0.8) {};

      \coordinate (W1) at (-1.6,-1.6) {};
      \coordinate (W2) at (-0.8,-0.8) {};
      \coordinate (W3) at (0.8, -0.8) {};
      \coordinate (W4) at (1.6, -1.6) {};
      
      \draw[Wilson_1] (WL)  -- (cusp) -- (WR);
      
      \draw (B1) -- (B2) -- (B3) -- (B1);
      \draw[Gluon_1] (B1) -- (W1) ;
      \draw[Gluon_1] (W4) -- (B2) ;
      \draw[Gluon_1] (W3) -- (B4) ;
      \draw[Gluon_1] (W2) -- (B4) ;
      \draw[Gluon_1] (B3) -- (B4) ;
      
\end{tikzpicture}

%% file: Plots/Dia_fRAA.tex
\begin{tikzpicture}[scale=0.8, baseline=(current bounding box.center)]

      \node at (0,0) {~};
      \node at (0,-2.1) {~};

      \coordinate (WL) at (-2,-2)   {};
      \coordinate (cusp) at (0,0)   {};
      \coordinate (WR) at (2,-2) {};
      
      \coordinate (B1) at (-0.5,-1.9) {};
      \coordinate (B2) at (+0.5,-1.9) {};
      \coordinate (B3) at (0,-1.4) {};

      \coordinate (W1) at (-1.2,-1.2) {};
      \coordinate (W2) at (-0.7,-0.7) {};
      \coordinate (W3) at (1.6, -1.6) {};
      \coordinate (W4) at (1.2, -1.2) {};
      
      \draw[Wilson_1] (WL)  -- (cusp) -- (WR);
      
      \draw (B1) -- (B2) -- (B3) -- (B1);
      \draw[Gluon_1] (B1) -- (W1) ;
      \draw[Gluon_1] (W4) -- (B2) ;
      \draw[Gluon_1] (B3) -- (W2) ;
      
      \draw[Gluon_1]  (W3) arc [start angle=-45, end angle=135, x radius=0.55, y radius=0.55];
      
\end{tikzpicture}

%% file: Plots/Dia_fdRdF.tex
\begin{tikzpicture}[scale=0.8, baseline=(current bounding box.center)]

      \node at (0,0) {~};
      \node at (0,-2.1) {~};

      \coordinate (WL) at (-2,-2)   {};
      \coordinate (cusp) at (0,0)   {};
      \coordinate (WR) at (2,-2) {};
      
      \coordinate (B1) at (-0.4,-1.1) {};
      \coordinate (B2) at (0.4,-1.1) {};
      \coordinate (B3) at (0.4,-1.9) {};
      \coordinate (B4) at (-0.4,-1.9) {};
      
      \coordinate (W1) at (-0.8,-0.8) {};
      \coordinate (W4) at (-1.257,-1.257) {};
      \coordinate (W2) at (0.8,-0.8) {};
      \coordinate (W3) at (1.257,-1.257) {};
      
      \draw[Wilson_1] (WL)  -- (cusp) -- (WR);
      \draw (B1) -- (B2) -- (B3) -- (B4) -- (B1);
      \draw[Gluon_1] (B1) -- (W1) ;
      \draw[Gluon_1] (B4) -- (W4) ;
      \draw[Gluon_1] (W3) -- (B3) ;
      \draw[Gluon_1] (W2) -- (B2) ;
\end{tikzpicture}

%% file: Plots/Dia_RAAA.tex
\begin{tikzpicture}[scale=0.8, baseline=(current bounding box.center)]

      \node at (0,0) {~};
      \node at (0,-2.1) {~};

      \coordinate (WL) at (-2,-2)   {};
      \coordinate (cusp) at (0,0)   {};
      \coordinate (WR) at (2,-2) {};
      
      \coordinate (B1) at (-0.5,-1.9) {};
      \coordinate (B2) at (+0.5,-1.9) {};
      \coordinate (B3) at (-0.3,-1.2) {};

      \coordinate (W1) at (-1.6,-1.6) {};
      \coordinate (W2) at (-0.5,-0.5) {};
      \coordinate (W3) at (0.5, -0.5) {};
      \coordinate (W4) at (1.1, -1.1) {};
      \coordinate (W5) at (1.6, -1.6) {};
      
      \draw[Wilson_1] (WL)  -- (cusp) -- (WR);
      
      \draw[Gluon_1] (B1) -- (W1) ;
      \draw[Gluon_1] (B2) -- (B1) ;
      \draw[Gluon_1] (W5) -- (B2) ;
      \draw[Gluon_1] (W2) -- (B3) ;
      \draw[Gluon_1] (B1) -- (B3) ;
      \draw[Gluon_1] (B2) .. controls (0.7,-1.4) .. (W4) ;
      \draw[Gluon_1] (W3) .. controls (0.3,-0.9) .. (B3) ;

\end{tikzpicture}

%% file: Plots/Dia_dRdA.tex
\begin{tikzpicture}[scale=0.8, baseline=(current bounding box.center)]
      
      \node at (0,0) {~};
      \node at (0,-2.1) {~};
      
      \coordinate (WL) at (-2,-2)   {};
      \coordinate (cusp) at (0,0)   {};
      \coordinate (WR) at (2,-2) {};
      
      \coordinate (W1) at (-1.6,-1.6) {};
      \coordinate (W2) at (-1.2,-1.2) {};
      \coordinate (W3) at (-0.7, -0.7) {};
      \coordinate (W4) at (-0.3, -0.3) {};
      \coordinate (W5) at (0.3, -0.3) {};
      \coordinate (W6) at (0.7, -0.7) {};
      \coordinate (W7) at (1.2,-1.2) {};
      \coordinate (W8) at (1.6,-1.6) {};
      
      \draw[Wilson_1] (WL)  -- (cusp) -- (WR);

      \draw[Gluon_1] (W6) .. controls (-0.3,-1.2) .. (W2) ;
      \draw[white,fill=white] (0,-1.05) circle (8pt);
      \draw[Gluon_1] (W7) .. controls (0.3,-1.2) .. (W3) ;
      \draw[white,fill=white] (0.01,-1.05) circle (6.1pt);
      \draw[Gluon_1] (W5) .. controls (-0.1,-1.4) .. (W1) ;
      \draw[white,fill=white] (-0.025,-1.045) circle (4pt);
      \draw[Gluon_1] (W8) .. controls (0.1,-1.4) .. (W4) ;
      
\end{tikzpicture}

%% file: Plots/Wilson_Cusp.tex

     \begin{tikzpicture}[scale=1.5]

	  \coordinate (cusp) at (0,0) {};
	  \coordinate (InW) at (-1,-1) {};
	  \coordinate (OutW) at (1,-1) {};
	  \coordinate (InPhi) at (0,-1) {};
	  \coordinate (le) at (-0.5,-0.5) {};
	  \coordinate (ri) at (+0.5,-0.5) {};
	  \coordinate (aux) at (+0.4,0.4) {};
	  
	  \draw[Wilson] (InW)  --  node[pos=0, anchor = north] {$v_1$} (le) -- node[pos=0, anchor = south east] {$~$} (cusp) -- node[pos=1, anchor = south west] {$~$} (ri) -- node[pos=1, anchor = north] {$v_2$} (OutW) {};	  
      
          \draw (cusp) -- (aux);
	  \draw (-45:0.35) arc [start angle=-45, end angle=45, radius=0.35];
	  \node (NN) at (0.2,0) {$\phi$};
      
      \end{tikzpicture}

%% file: Plots/Topo_1_box.tex
\begin{tikzpicture}[scale=1.1, baseline=(current bounding box.north)]
  
      \node at (-0.6,1.5) {(1)};
      \node at (-0.6,-0.65) {~};
      
      \coordinate (W0) at (-0.5,0)   {};
      \coordinate (W1) at (0,0) {};
      \coordinate (W2) at (1,0) {};
      \coordinate (W3) at (2,0) {};
      \coordinate (W4) at (3,0) {};
      \coordinate (W5) at (3.5,0)   {};
      \coordinate (P1) at (0.5,1.5)   {};
      \coordinate (P2) at (1,1)   {};
      \coordinate (P3) at (2,1)   {};
      \coordinate (P4) at (2.5,1.5)   {};

      \draw[Wilson_arrow] (W0)  -- node[pos=0.5, anchor = south] {\footnotesize{$v$}} (W1);
      \draw[Wilson_blank] (W1) -- node[pos=0.5, anchor = north] {\footnotesize{~}}  (W2) --  node[pos=0.5, anchor = north] {\footnotesize{~}}  (W3) --  node[pos=0.5, anchor = north] {\footnotesize{~}}  (W4) -- (W5) {};
      \draw[QuadProp] (W1) -- node[pos=0.5, anchor = east] {\footnotesize{~}} (P1) {};
      \draw[QuadProp] (W2) -- node[pos=0.5, anchor = east] {\footnotesize{~}} (P2) {};
      \draw[QuadProp] (W3) -- node[pos=0.5, anchor = east] {\footnotesize{~}} (P3) {};
      \draw[QuadProp] (W4) -- node[pos=0.5, anchor = east] {\footnotesize{~}}(P4) {};
      
      \draw[QuadProp] (P1) -- node[pos=0.5, anchor = south] {\footnotesize{~}}(P2) --  node[pos=0.5, anchor = south] {\footnotesize{~}}(P3) -- node[pos=0.5, anchor = south] {\footnotesize{~}}(P4) -- node[pos=0.5, anchor = south] {\footnotesize{~}}(P1) {};

\end{tikzpicture}

%% file: Plots/Topo_2_box.tex
\begin{tikzpicture}[scale=1.1, baseline=(current bounding box.north)]

      \node at (-0.6,1.5) {(2)};
      \node at (-0.6,-0.65) {~};
      
      \coordinate (W0) at (-0.5,0)   {};
      \coordinate (W1) at (0,0) {};
      \coordinate (W2) at (1,0) {};
      \coordinate (W3) at (2,0) {};
      \coordinate (W4) at (3,0) {};
      \coordinate (W5) at (3.5,0)   {};
      \coordinate (P1) at (0.5,1.5)   {};
      \coordinate (P2) at (1,1)   {};
      \coordinate (P3) at (2,1)   {};
      \coordinate (P4) at (2.5,1.5)   {};
      
      \draw[Wilson_arrow] (W0)  -- node[pos=0.5, anchor = south] {\footnotesize{$v$}} (W1);
      \draw[Wilson_blank] (W1) -- node[pos=0.5, anchor = north] {\footnotesize{~}}  (W2) --  node[pos=0.5, anchor = north] {\footnotesize{~}}  (W3) --  node[pos=0.5, anchor = north] {\footnotesize{~}}  (W4) -- (W5) {};
      \draw[QuadProp] (W1) -- node[pos=0.5, anchor = east] {\footnotesize{~}} (P1) {};
      \draw[QuadProp] (W2) -- node[pos=0.5, anchor = east] {\footnotesize{~}} (P2) {};
      \draw[QuadProp] (W3) .. controls (3,0.8) .. node[pos=0.5, anchor = east] {\footnotesize{~}} (P4) {};
      \draw[white,fill=white] (2.55,0.45) circle (0.12);
      \draw[QuadProp] (W4) -- node[pos=0.5, anchor = east] {\footnotesize{~}}(P3) {};
      
      \draw[QuadProp] (P1) -- node[pos=0.5, anchor = south] {\footnotesize{~}}(P2) --  node[pos=0.5, anchor = south] {\footnotesize{~}}(P3) -- node[pos=0.5, anchor = south] {\footnotesize{~}}(P4) -- node[pos=0.5, anchor = south] {\footnotesize{~}}(P1) {};
      
\end{tikzpicture}

%% file: Plots/Topo_3_box.tex
 \begin{tikzpicture}[scale=1.1, baseline=(current bounding box.north)]
      \node at (-0.6,1.5) {(3)};
      \node at (-0.6,-0.65) {~};
      
      \coordinate (W0) at (-0.5,0)   {};
      \coordinate (W1) at (0,0) {};
      \coordinate (W2) at (1,0) {};
      \coordinate (W3) at (2,0) {};
      \coordinate (W4) at (3,0) {};
      \coordinate (W5) at (3.5,0)   {};
      \coordinate (P1) at (0.5,1.5)   {};
      \coordinate (P2) at (1,1)   {};
      \coordinate (P3) at (2,1)   {};
      \coordinate (P4) at (2.5,1.5)   {};
      
      \draw[Wilson_arrow] (W0)  -- node[pos=0.5, anchor = south] {\footnotesize{$v$}} (W1);
      \draw[Wilson_blank] (W1) -- node[pos=0.5, anchor = north] {\footnotesize{~}}  (W2) --  node[pos=0.5, anchor = north] {\footnotesize{~}}  (W3) --  node[pos=0.5, anchor = north] {\footnotesize{~}}  (W4) -- (W5) {};
      \draw[QuadProp] (W1) -- node[pos=0.5, anchor = east] {\footnotesize{~}} (P1) {};
      \draw[QuadProp] (W2) -- node[pos=0.5, anchor = east] {\footnotesize{~}} (P3) {};
      \draw[white,fill=white] (1.5,0.5) circle (0.12);
      \draw[QuadProp] (W3) -- node[pos=0.5, anchor = east] {\footnotesize{~}} (P2) {};
      \draw[QuadProp] (W4) -- node[pos=0.5, anchor = east] {\footnotesize{~}}(P4) {};
      
      \draw[QuadProp] (P1) -- node[pos=0.5, anchor = south] {\footnotesize{~}}(P2) --  node[pos=0.5, anchor = south] {\footnotesize{~}}(P3) -- node[pos=0.5, anchor = south] {\footnotesize{~}}(P4) -- node[pos=0.5, anchor = south] {\footnotesize{~}}(P1) {};
      
\end{tikzpicture}

%% file: Plots/Topo_4.tex
 \begin{tikzpicture}[scale=1.4, baseline=(current bounding box.north)]
  
      \node at (-0.5,1) {(4)};
      \node at (-0.5,-0.65) {~};
      
      \coordinate (W0) at (-0.5,0)   {};
      \coordinate (W1) at (0,0) {};
      \coordinate (W2) at (1,0) {};
      \coordinate (W3) at (2,0) {};
      \coordinate (W4) at (3,0) {};
      \coordinate (W5) at (4,0) {};
      \coordinate (W6) at (4.5,0)   {};
      \coordinate (P1) at (1,1)   {};
      \coordinate (P2) at (2,1)   {};
      \coordinate (P3) at (3,1)   {};
      
      \draw[Wilson_arrow] (W0)  -- node[pos=0.5, anchor = south] {\footnotesize{$v$}} (W1);
      \draw[Wilson_blank] (W1) -- node[pos=0.5, anchor = north] {\footnotesize{1}}  (W2) --  node[pos=0.5, anchor = north] {\footnotesize{2}}  (W3) --  node[pos=0.5, anchor = north] {\footnotesize{4}} (W4) --  node[pos=0.5, anchor = north] {\footnotesize{3}}  (W5) -- (W6) {};
      \draw[QuadProp] (W1) -- node[pos=0.5, anchor = south east] {\footnotesize{5}} (P1) -- node[pos=0.5, anchor = south] {\footnotesize{6}} (P2) -- node[pos=0.5, anchor = south] {\footnotesize{8}} (P3) -- node[pos=0.5, anchor = south west] {\footnotesize{7}} (W5) {};
      \draw[QuadProp] (W2) -- node[pos=0.5, anchor = east] {\footnotesize{9}} (P1) {};
      \draw[QuadProp] (W3) -- node[pos=0.5, anchor = east] {\footnotesize{13}} (P2) {};
      \draw[QuadProp] (W4) -- node[pos=0.5, anchor = east] {\footnotesize{14}}(P3) {};
\end{tikzpicture}

%% file: Plots/Topo_5.tex
\begin{tikzpicture}[scale=1.4, baseline=(current bounding box.north)]
  
      \node at (-1.25,1) {(5)};
      \node at (-0.5,-0.65) {~};
      
      \coordinate (W0) at (-1.25,0)   {};
      \coordinate (W1) at (-0.75,0) {};
      \coordinate (W2) at (0,0) {};
      \coordinate (W3) at (1.5,0) {};
      \coordinate (W4) at (3,0) {};
      \coordinate (W5) at (3.5,0) {};
      \coordinate (P1) at (0.9,1)   {};
      \coordinate (P2) at (1.5,0.5)   {};
      \coordinate (P3) at (2.1,1)   {};
      
      \draw[Wilson_arrow] (W0)  -- node[pos=0.5, anchor = south] {\footnotesize{$v$}} (W1);
      \draw[Wilson_blank] (W1) -- node[pos=0.5, anchor = north] {\footnotesize{15}}  (W2) --  node[pos=0.5, anchor = north] {\footnotesize{16}}  (W3) --  node[pos=0.5, anchor = north] {\footnotesize{2}} (W4) -- (W5)  {};
      \draw[QuadProp] (W2) -- node[pos=0.5, anchor = south east] {\footnotesize{12}} (P1) -- node[pos=0.5, anchor = south] {\footnotesize{13}} (P3) -- node[pos=0.5, anchor = south west] {\footnotesize{6}} (W4) {};
     \draw[QuadProp] (W3) -- node[pos=0.5, anchor = east] {\footnotesize{7}} (P2) {};
      \draw[QuadProp] (P1) -- node[pos=0.5, anchor = north east] {\footnotesize{14}} (P2) -- node[pos=0.5, anchor = north west] {\footnotesize{8}} (P3) {};
      \draw[QuadProp] (W1) arc [start angle=180, end angle=360, x radius=1.125, y radius=0.6] node[pos=0.5, anchor = south] {\footnotesize{9}};
      
\end{tikzpicture}